\documentclass[reprint,showpacs,amsmath,amssymb,aps,pra]{revtex4-1}

\usepackage{footmisc}
\usepackage{graphicx}
\usepackage{dcolumn}
\usepackage{bm}
\begin{document}

\preprint{APS/123-QED}

\title{Quantum Traversal Time Across a Potential Well}

\author{Dean Alvin L. Pablico}
 \altaffiliation[]{dlpablico@up.edu.ph}
\author{Eric A. Galapon}
 \email{eagalapon@up.edu.ph}
\affiliation{
Theoretical Physics Group, National Institute of Physics, University of the Philippines, Diliman Quezon City, 1101 Philippines
}

\date{\today}

\begin{abstract}
We consider the quantum traversal time of an incident wave packet across a potential well using the theory of quantum time of arrival (TOA)-operators. This is done by constructing the corresponding  TOA-operator across a potential well via quantization. The expectation value of the potential well TOA-operator is compared to the free particle case for the same incident wave packet. The comparison yields a closed-form expression of the quantum well traversal time which explicitly shows the classical contributions of the positive and negative momentum components of the incident wave packet and a purely quantum mechanical contribution significantly dependent on the well depth. An incident Gaussian wave packet is then used as an example. It is shown that for shallow potential wells, the quantum well traversal time approaches the classical traversal time across the well region when the incident wave packet is spatially broad and approaches the expected quantum free particle traversal time when the wave packet is localized. For deep potential wells, the quantum traversal time oscillates from positive to negative implying that the wave packet can be advanced or delayed.  
\end{abstract}


\maketitle

\section{\label{sec:level1}Introduction}

One of the simplest and most studied potentials in quantum mechanics are rectangular one-dimensional potentials such as the potential barrier and well. These type of potentials offer interesting quantum mechanical phenomena such as the quantum tunneling through a potential barrier  and quantum reflection and transmission through a potential well. Such phenomena are all predicted by solving the time-independent Schr\"{o}dinger equation. However, things get complicated when one has to incorporate the time-dependent picture of these quantum mechanical effects, for example, the traversal time, which is the time it takes for a quantum wave packet or particle to traverse a given region of space. In classical mechanics, one can just use a stopwatch to find for the traversal time of a classical particle. However, the concept of traversal time in the context of quantum mechanics has been controversial to this date since standard quantum theory does not offer clear definition and unique treatment of quantum traversal time \cite{muga,muga2}. 

The most famous quantum traversal time problem is the quantum tunneling time problem, the problem of how long a particle tunnels through a potential barrier which is classically forbidden \cite{tunnel1,tunnel2}. It is almost as old as quantum mechanics itself \cite{maccoll} and has attracted much attention with diverse and contradicting opinions \cite{tunnel1,keller2}. 
Initial studies on quantum tunneling using attosecond angular streaking technique, which can time the release of electrons in strong-field ionization with a precision of a few attoseconds, confirm that tunneling happens instantaneously \cite{keller,keller2}. In contrast, a study on multi-electron atoms claim evidence for finite tunnelling times \cite{camus}. However, a more recent paper by Sainadh et al. reported an instantaneous tunneling in atomic hydrogen using attoclock and momentum-space imaging \cite{sainadh}. Although there is still no consensus on whether or not quantum tunneling occurs instantaneously or not, the attosecond science community is more inclined on instantaneous quantum tunnelling time \cite{zhao,keller2}. Such experimental results are also consistent with the independent theoretical predictions of Galapon \cite{galapon1} and Pollak et al. \cite{pollak,pollak2,pollak3}. 

Now that the tunneling time for particles passing through a potential barrier may be instantaneous, what would the corresponding quantum traversal time be when particles pass through a potential well instead? This may sound simple at first since classical mechanics already tells us that the particle would speed up in the well region because its incident energy is increased by the potential depth. This implies that the traversal time in the well region is always lesser compared to the traversal time in the free region. However, this is not necessarily true if one employs a quantum mechanical treatment of traversal time. 

Li and Wang have reported a negative phase time for particles passing through a potential well \cite{Li}. Their result implies that quantum particles pass through the well region at a negative phase velocity. This further suggests that the quantum particles seem to leave the potential well before entering it. The existence of the negative phase time has been experimentally verified by Vetter et al. using electromagnetic wells realized by waveguides filled with different dielectrics \cite{Vetter}. On the other hand, Chen and Li have also shown that the group delay for Dirac particles travelling through a potential well can be negative under specific conditions \cite{Chenli}. Another paper by Chen and Li discussed the mechanism of superluminal traversal time from the viewpoint of interference between multiple finite wave packets, due to multiple reflections inside a potential well or barrier \cite{Chenli2}. Furthermore, Los et al. has showed that the particle arrival time and dwell time in a potential barrier or well is dependent on the positive and negative momentum components, and their inteference of an incident wave packet \cite{Los}. Finally, Muga et al. have shown that potential square wells lead to much larger time advancements than square barriers \cite{delgado}. 

We are then motivated to consider this problem anew using the theory of quantum time of arrival (TOA) operators proposed in Refs. \cite{galapon2supra,Magadan}. The basic idea is to incorporate time and time quantities as dynamical observables, like any other observables in standard quantum mechanics. For our case, the quantum well traversal time for quantum particles is determined under the hypothesis that we can meaningfully construct a TOA-operator $\hat{T}$ corresponding to an arrival at some point  $q$ in our configuration space for a given interaction potential $V(q)$. This  TOA-operator is then constucted via quantization of the classical TOA \cite{Magadan}. The expectation values of the potential well and free particle  TOA-operators are then compared for the same incident wave packet.

It is shown in this paper that the quantum traversal time across the well region is dependent on the initial state of the incident wave packet $\tilde{\Psi}(k)$ in momentum space representaton and the width and depth of the potential well. It is also expressed as the weighted sum of the classical traversal times on top of the well region with weights $|\tilde{\Psi}(k)|^2$ and $|\tilde{\Psi}(-k)|^2$ and the traversal time inside of the potential well with a non-positive definite weight $\mbox{Im}\left[2\,\tilde{\Psi}(ik)\tilde{\Psi}^*(-ik)\right]$.  An incident Gaussian wave packet is then used as an example. It is found that for shallow potential wells, the expected quantum well traversal time approaches the classical traversal time across the well region when the incident wave packet is spatially broad. Meanwhile, it approaches the free particle quantum traversal time for localized or spatially narrow wave packets. For deep potential wells, the quantum well traversal time oscillates from positive to negative implying that the wave packet, on average, can be advanced or delayed.  

This paper provides concrete testable predictions that may clarify the nature of time as a quantum dynamical observable. More specifically, it gives deeper understanding and new insights about the proper treatment of quantum traversal time for quantum systems since standard quantum theory does not give us a clear quantum mechanical generalization of the classical traversal time. Furthermore, it has potential application in electronic devices since the concept of group time delay, an aspect of traversal time, is important in quantum particles’ transport in various semiconductor devices \cite{ban,wang}.   

The rest of the paper is organized as follows. The  TOA-operator is constructed via quantization in Sec. \ref{sec:level2}. The quantum traversal time across the potential well is determined by comparing the  expectation values of the  TOA-operators in the presence and absence of the potential well in Sec. \ref{sec:level3}. The relation between the quantum well and barrier traversal times is investigated in Sec. \ref{sec:level4}. An incident Gaussian wave packet is used as an example to explicitly calculate the quantum traversal time across the potential well in Sec. \ref{sec:level4}. Lastly, the summary and conclusions of the paper is given in Sec. \ref{sec:level5}.

\section{\label{sec:level2}Quantum Time of Arrival (TOA)-Operator}

\begin{figure}
	\includegraphics[width=0.48\textwidth]{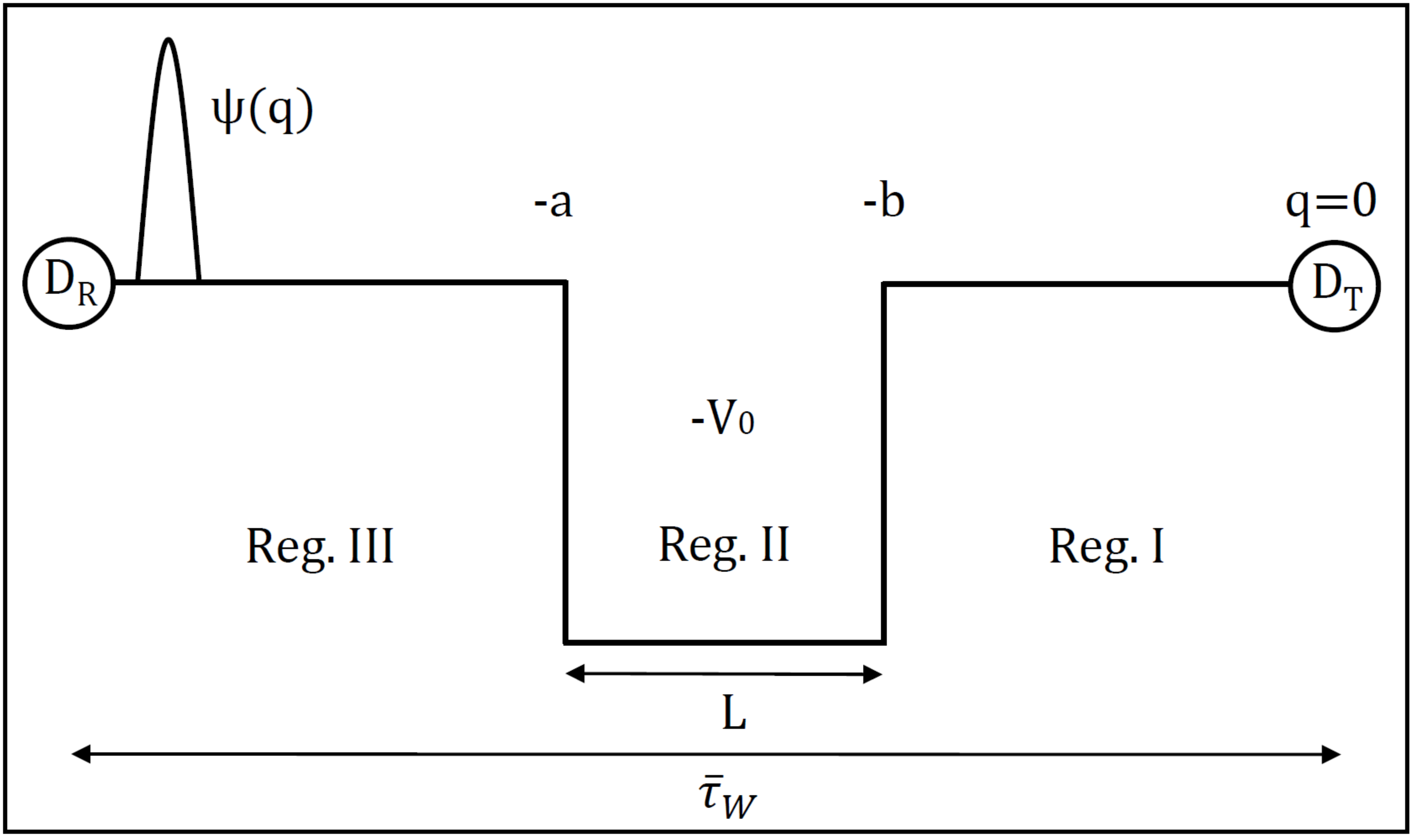}
	\caption{Measurement scheme in the presence of a potential well. The potential well $V(q)=-V_0$ of length $L$ is located at $-a<q<-b$ with the arrival point at $q=0$. The corresponding average TOA $\bar{\tau}_W$ at $D_T$ is computed.}
	\label{fig:measurewell}
\end{figure}
\begin{figure}
	\includegraphics[width=0.48\textwidth]{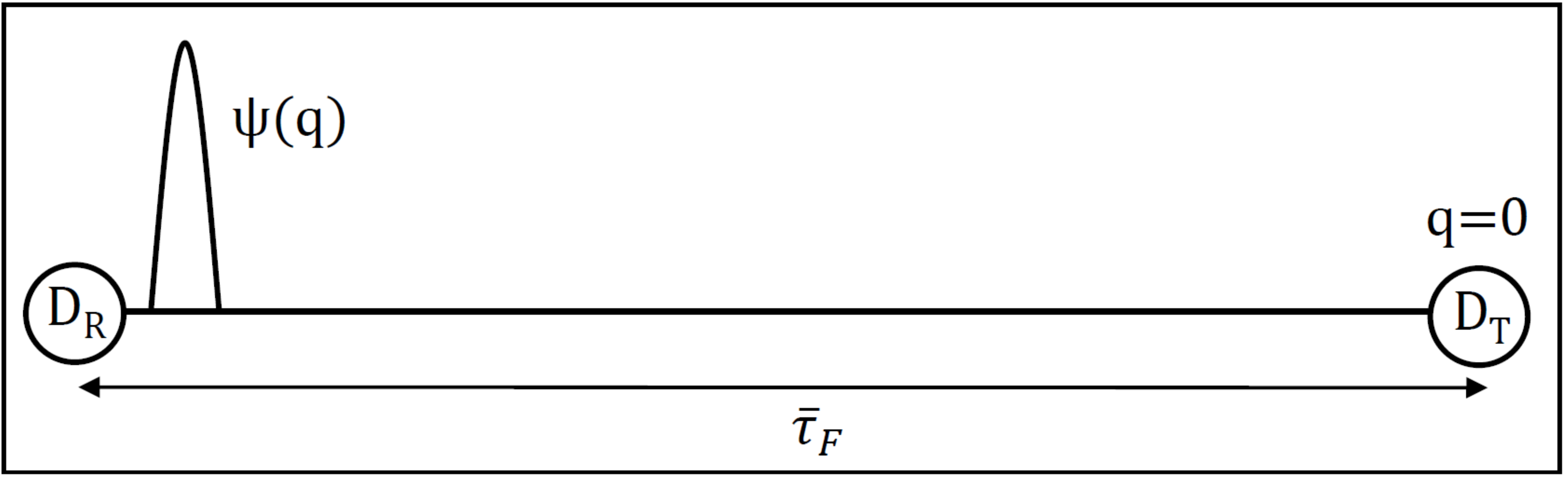}
	\caption{Measurement scheme in the abscence of a potential well. The corresponding average TOA $\bar{\tau}_F$ at $D_T$ is computed.}
	\label{fig:measurefree}
\end{figure}

The same operational definition of the quantum traversal time prescribed in Refs. \cite{galapon1,sombillodouble,sombillotecr} is used. The measurement schemes in the presence and absence of the potential well are shown in Figs. (\ref{fig:measurewell}) and (\ref{fig:measurefree}), respectively and are described as follows. A detector $D_T$ is placed at the origin to announce the arrival of a particle and a detector $D_R$ at the far left of $D_T$. A potential well $V(q)=-V_0$ of width $L$ is situated in between $D_T$ and $D_R$ at $-a<q<-b$ where $a$ and $b$ are both positive. A wave packet $\psi(q)$ is prepared between $D_R$ and the potential well and launched at time \textit{t}=0. The time of arrival of the particle at the origin is recorded when $D_T$ clicks; if not, no data is collected when $D_R$ clicks. This is repeated  a large number of times, with $\psi(q)$ as the initial state for every repeat, then the average time of arrival $\bar{\tau}_W$ at $D_T$ is computed. A similar experiment for a free particle is performed, that is in the absence of the potential well, as shown in Fig. (\ref{fig:measurefree}). The average free time of arrival $\bar{\tau}_F$ at $D_T$ is then computed from the new time of arrival data. A comparison between the average TOAs $\bar{\tau}_W$ and $\bar{\tau}_F$ is made. The expected traversal time across the potential well is deduced from the TOA difference given by
\begin{equation}\label{first}
\Delta\tau=\bar{\tau}_F-\bar{\tau}_W.
\end{equation}
Equation (\ref{first}) implies three possible cases, i.e., $ \Delta\tau$ is 0, positive and negative. The first case ($\Delta\tau=0$) suggests that the average traversal time for a free particle case and in the presence of the well is the same. The case $\Delta\tau>0$ suggests that the quantum particle, on average, passed through the potential well region earlier than the free particle case. Meanwhile, the case $\Delta\tau<0$ implies that the free particle, on average, is advanced so that the quantum particle travelling through the potential well is delayed. The measurement scheme for each case is chosen because it avoids altering the propagation of the incident wave packet and hence, providing an indirect but realistic and accurate way of obtaining the well traversal time. Note, however, that Eq. (\ref{first}) is not the well traversal time itself but the traversal time can be deduced from it.

We now hypothesize that this measurement scheme can be theoretically described by  TOA-operators. Let $\hat{T}_W$ be the  TOA-operator in the presence of the potential well, and $\hat{T}_F$ be the  TOA-operator in the absence of the well. The operator $\hat{T}_F$ is defined as the free particle  TOA-operator, which is the quantization of its classical time of arrival \cite{galapon2supra,galaponsize}. The operator $\hat{T}_W$ is constructed by quantization using the theory of quantum time of arrival in the presence of an interaction potential $V(q)$ described in Refs. \cite{galapon2supra,Magadan}. We identify $\bar{\tau}_W$ as the expectation value $\bar{\tau}_W$ = $\langle\psi| \hat{T}_W|\psi\rangle$ and $\bar{\tau}_F$ as the expectation value $\bar{\tau}_F$ = $\langle\psi| \hat{T}_F|\psi\rangle$ of the two  TOA-operators $\hat{T}_W$ and $\hat{T}_F$, respectively. It was already shown by one of us that the expectation value $\bar{\tau}_F$ leads to the correct classical value, where the classical TOA exists, in the limit as $\hbar$ approaches zero \cite{galaponsize}. It will be shown later that the expectation value $\bar{\tau}_W$ also leads to its correct classical limit. Hence, the time of arrival difference given in Eq. (\ref{first}) becomes
\begin{equation}\label{second}
\Delta\tau=\langle\psi| \hat{T}_F|\psi\rangle  - \langle\psi| \hat{T}_W|\psi\rangle.
\end{equation}

For analytic or piecewise constant potentials, a  TOA-operator can be constructed by quantization of the classical TOA \cite{galapon2supra,Magadan}. In coordinate representation, the quantized  TOA-operator for arrival at the origin is the integral operator 
\begin{equation}\label{third}
(\hat{T}\phi)(q)=\int_{-\infty}^{\infty}\frac{\mu}{\textit{i}\hbar}T(q,q')\mbox{sgn}(q-q')\phi(q')dq',
\end{equation}
where $\mu$ is the mass of the particle, $\mbox{sgn}(x)$ is the sign function and $T(q,q')$ is the time kernel factor for a chosen quantization. For Weyl quantization, the time kernel is given by, 
\begin{equation}\label{fourth}
T(q,q')=\frac{1}{2}\int_{0}^{\eta} \,\, _0F_1\bigg(;1;\frac{\mu}{2\hbar^2}\zeta^2{V(\eta)-V(\eta')}\bigg) ds,
\end{equation}
in which $_0F_1(;1;z)$ is a specific hypergeometric function, $\zeta =q-q'$ , and $\eta = (q+q')/2$ \cite{galapon2supra}. In general, the time kernels via different quantizations are not the same due to the non-commutativity of the position and momentum operators. \cite{Magadan}. However, it can be easily shown that for the case of constant piecewise potentials, such as that of potential barriers and wells, the time kernels are equal for Weyl, Born-Jordan and simple symmetric quantizations. 

For the free particle case, substituting $V(q)=0$ into Eq. (\ref{fourth}) and using the identity $_0\textit{F\textsubscript{1}}(;1;0)=1$ yields the time kernel factor given by
\begin{equation}\label{fifth}
\textit{T}_F(q,q') = \frac{q+q'}{4}. 
\end{equation}
Substituting $T_F(q,q')$ back into Eq. (\ref{third}) gives the free  TOA-operator in its integral form \cite{galaponsize}.

The  TOA-operator $\hat{T}_W$ across the potential well is constructed by solving first for the time kernels using Eq. (\ref{fourth}) with reference to Fig. (\ref{fig:measurewell}). The potential \textit{V(q)} in configuration space is mapped into the same potential in the $\eta$ coordinate such that the arrival point is at $\eta=0$. A change of variables is done using the relations $\eta=(q+q')/2$ and $\zeta=q-q'$ so that the time kernel factor given in Eq. (\ref{fourth}) assumes the form $T(q,q')=\tilde{T}(\eta,\zeta)$. The time kernel factor $\tilde{T}(\eta,\zeta)$ is obtained by dividing the $\eta$ coordinate into three non-overlapping regions separated by the edges of the well as shown in Fig. (\ref{fig:measurewell}). Thus, $\tilde{T}(\eta,\zeta)$ has three pieces corresponding to the three regions where $\eta$ may fall which are given by
\begin{equation}\label{kernel}
\begin{aligned}
&\tilde{T}_1(\eta,\zeta) = \frac{\eta}{2} , \\
&\tilde{T}_2(\eta,\zeta) = \frac{\eta}{2}-\frac{b}{2} \big[J_0(\kappa|\zeta|)-1\big] ,\\
&\tilde{T}_3(\eta,\zeta) = \frac{\eta}{2}-\frac{L}{2} \big[I_0(\kappa|\zeta|)-1\big] ,
\end{aligned}
\end{equation}
where $\kappa=\sqrt{2 \mu V_0}/\hbar$, and $L=a-b$ is the width of the potential well.

We now prove that Eq. (\ref{third}) gives the quantization of the classical time of arrival across the well by showing that the constructed  TOA-operator gives the correct classical limit. The limit is obtained by taking the inverse Weyl-Wigner transform of the kernel $\langle q|\hat{T}|q' \rangle$ given by
\begin{equation}\label{sixth}
t_n(q_0,p_0)=\frac{\mu}{i\hbar}\int_{-\infty}^{\infty}\tilde{T}_n(q_0,\zeta)\mbox{sgn}(\zeta) e^{ip\textsubscript{o}\zeta/\hbar} d\zeta,
\end{equation}
where $q_0$ and $p_0$ are the initial position and momentum, respectively.  The integral in Eq. (\ref{sixth}) is to be understood in the distributional sense. The subsrcript $n$ represents the region where the initial position $q_0$ lies in the three possible regions. For example, the time kernel factor $\tilde{T}_1(q_0,\zeta)$ gives the classical time of arrival $t_1$ in which the particle started somewhere within region I.

We substitute $\tilde{T}_1(q_0,\zeta)$, $\tilde{T}_2(q_0,\zeta)$, and $\tilde{T}_3(q_0,\zeta)$ into Eq. (\ref{sixth}). Integrals involving Bessel functions are evaluated by expanding them in their power series representations, exchanging the order of summation and integration and then performing a term by term integration using the integral identity 
\begin{equation}\label{seventh}
\int_{-\infty}^{\infty} \nu^{m-1} \mbox{sgn} (\nu) e^{-ix\nu} d\nu = \frac{2(m-1)!}{i^m x^m},
\end{equation}
(the inverse Fourier transform of \cite{shilov}, p.360, no.18) to obtain the classical limit. We then sum the resulting series which leads to the following classical limits corresponding to the three possible locations of $q_0$,
\begin{equation}
\begin{aligned}
&t\textsubscript{1}(q_0,p_0) = -\mu\frac{q_0}{p_0} , \\
&t_2(q_0,p_0) = -\mu\frac{q_0+b}{p_0}+\frac{\mu b}{\sqrt{p_0^2 - 2\mu V_0}},\\
&t\textsubscript{3}(q_0,p_0) = -\mu\frac{q_0+L}{p_0}+\frac{\mu L}{\sqrt{p_0^2 + 2\mu V_0}},
\end{aligned}
\end{equation}
provided that $2 \mu V _0/p_0^2 < 1$.

Notice that $t\textsubscript{1}$ is just the expected classical time of arrival in the free region of length $q_0$. For $t\textsubscript{2}$, the first term is the traversal time on top of the well region of length $-(q_0+b)>0$ with momentum $p_0$ and the second term is the traversal time across the free segment \textit{b} with momentum $\sqrt{p_0^2-2 \mu V_0}$. On the other hand, the first term of $t\textsubscript{3}$ is the traversal time before and after the potential well with momentum $p_0$. Note that the second term of $t\textsubscript{3}$ is just the known classical traversal time across the potential well,
\begin{equation}\label{eigth}
t_{classical}=\frac{\mu L}{\sqrt{p_0^2 + 2\mu V_0}}.
\end{equation}
Hence, the constructed  TOA-operator reduces to the correct classical TOA expression in the classical limit.

\section{\label{sec:level3}Quantum well traversal Time}
The expectation value for a  TOA-operator $\hat{T}$ for a given incident wave packet $\psi(q)$  is given by
\begin{equation}\label{ninth}
\langle\psi|\hat{T}|\psi\rangle=\int_{-\infty}^{\infty}\int_{-\infty}^{\infty}\bar{\psi}(q)\psi(q')\frac{\mu}{\textit{i}\hbar}T(q,q')\mbox{sgn}(q-q')dq'dq.
\end{equation}
Let the incident wave packet takes the form $\psi(q)=\varphi(q)e^{ik_0q}$ with a momentum expectation value $\hbar k_0$ and group velocity $\hbar k_0/\mu$. Substituting $\psi(q)$ into Eq. (\ref{ninth}) and changing variables to $(\zeta,\eta)$, the expectation value evaluates to $\langle\psi|\hat{T}|\psi \rangle=\mbox{Im}(\tau^*)$ where $\tau^*$ is the complex-expected TOA given by
\begin{equation}\label{tenth}
\tau^*=-2\frac{\mu}{\hbar}\int_{-\infty}^{\infty}\int_{-\infty}^{\infty}\bar{\varphi}\Big(\eta-\frac{\zeta}{2}\Big)\varphi\Big(\eta+\frac{\zeta}{2}\Big)\tilde{T}(\eta,\zeta)e^{i k _0\zeta}d\eta d\zeta.
\end{equation}

We assume that the incident wave packet is infinitely differentiable and with support to the left of the well. The same assumption was used independently by Galapon \cite{galapon1} and Pollak \cite{pollak2} for the potential barrier case. Now, in the absence of the potential well, substituting the time kernel $\tilde{T}_F(\eta,\zeta)=\eta/2$ into Eq. (\ref{tenth}) gives the complex-expected TOA for the free particle case given by
\begin{equation}\label{eleventh}
\tau_F^*=-\frac{\mu}{\hbar}\int_{0}^{\infty}\int_{-\infty}^{\infty}\bar{\varphi}\Big(\eta-\frac{\zeta}{2}\Big)\varphi\Big(\eta+\frac{\zeta}{2}\Big)\eta e^{i k _0\zeta}d\eta d\zeta.
\end{equation}
In the presence of the well, we use the time kernel $\tilde{T}\textsubscript{3}(\eta,\zeta)$ only since we have assumed that the support of $\varphi(q)$ does not extend inside the well region. Substituting $\tilde{T}\textsubscript{3}(\eta,\zeta)$ into Eq. (\ref{tenth}) gives the complex-expected TOA in the presence of the potential well,
\begin{equation}\label{twelfth}
\begin{split}
\tau_W^*&= -\frac{\mu}{\hbar}\int_{0}^{\infty}\int_{-\infty}^{\infty}\bar{\varphi}\Big(\eta-\frac{\zeta}{2}\Big)\varphi\Big(\eta+\frac{\zeta}{2}\Big)(\eta + L) e^{i k _0\zeta}d\eta d\zeta \\
&+ \frac{\mu L}{\hbar}\int_{0}^{\infty}\int_{-\infty}^{\infty}\bar{\varphi}\Big(\eta-\frac{\zeta}{2}\Big)\varphi\Big(\eta+\frac{\zeta}{2}\Big)I_0(\kappa \zeta) e^{i k _0\zeta} d\eta d\zeta.
\end{split}
\end{equation}

Recall that the direct measurable quantity for deducing the well traversal time is the TOA difference given by Eq. (\ref{second}). In terms of the complex-expected TOAs $\tau_W^*$ and $\tau_F^*$, the TOA difference reduces to $\Delta \tau=$ Im($\Delta \tau^*$) = Im($\tau_F^*-\tau_W^*$). Using Eqs. (\ref{eleventh}) and (\ref{twelfth}), the complex-expected TOA difference leads to
\begin{equation}\label{thirteenth}
\Delta\tau^*=\frac{L}{v_0}(Q^*-R^*),
\end{equation}
where $v_0=\hbar k_0/\mu$ is the group velocity. The quantities $Q^*$ and $R^*$ are defined as follows,
\begin{equation}\label{14th}
Q^*= k_0\int_{0}^{\infty}d\zeta\, \Phi(\zeta)e^{i k_0\zeta},
\end{equation}
\begin{equation}\label{15th}
R^*= k_0\int_{0}^{\infty}d\zeta\, \Phi(\zeta)I_0(\kappa\zeta)e^{i k_0\zeta},
\end{equation}
where $\Phi(\zeta)$ is given by
\begin{equation}\label{16th}
\Phi(\zeta)= \int_{-\infty}^{\infty}d\eta \, \bar{\varphi}\left(\eta-\frac{\zeta}{2}\right)\varphi\left(\eta+\frac{\zeta}{2}\right).
\end{equation}

We understand first the underlying physical contents of the quantities $(L/v_0)Q$ and $(L/v_0)R$, where $Q= \mbox{Im}(Q^*)$ and $R= \mbox{Im}(R^*)$. This is done by investigating their corresponding asymptotic form in the high energy limit $k_0 \to \infty$ for fixed $\kappa$, i.e., fixed potential depth $V_0$. It was already shown by one of us that $Q\sim 1$ as $k_0 \to \infty$ \cite{galapon1}. This implies that the quantity  $(L/v_0)Q \sim L/v_0$ in the classical limit, which is just the classical traversal time for a quantum particle across a free region of length \textit{L} with speed $v_0$. We can therefore identify $\tau_F=(L/v_0)Q$ as the expected quantum traversal time for the free particle across the free region of length $L$.

For the quantity $(L/v_0)R$, notice that $R^*$ in Eq. (\ref{15th}) is a Fourier integral with respect to the asymptotic parameter $k_0$. By repeated integration by parts, collecting equal powers of $\hbar$, substituting $\kappa=\sqrt{2\mu V_0}/\hbar$ and  $k_0=\sqrt{2\mu E_0}/\hbar$, and taking the imaginary part lead us to the asymptotic expansion 
\begin{equation}\label{17th}
R \sim \sum_{j,l=0}^{\infty} \frac{(-1)^{j+l}}{2^{2l}}\frac{\hbar^{2j}}{(2\mu E_0)^j }\binom{2j+2l}{2l}\binom{2l}{l}\Big(\frac{V_0}{E_0}\Big)^l \Phi^{(2j)}(0)
\end{equation}
In the classical limit, only $j=0$ terms contribute since terms with the factor $\hbar$ must vanish. Using the normalization condition $\Phi(0)=1$, one finds
\begin{equation}\label{18th}
R\sim \sqrt{\frac{E_0}{E_0+V_0}}
\end{equation}
in the high energy limit $k_0 \to \infty$. It can be seen that Eq. (\ref{18th}) is just the square root of the ratio of the incident energy and the energy on top of the potential well. This can be written as the ratio of the free velocity $v_0$ and the velocity on top of the well $v_W$ such that $R\sim v_0/v_W$. In the classical limit, we then find $(L/v_0)R \sim L/v_W$ which is simply the classical traversal time across the potential well, consistent with Eq. (\ref{eigth}). Therefore, the quantity
\begin{equation}\label{tauw}
\tau_W= \frac{L}{v_0}R
\end{equation}
can be identified as the quantum traversal time across the potential well. This result can be related to optics. One can see that the classical limit $R\sim v_0/v_W$ is just the effective index of refraction of the well with respect to the incident wave packet, \textit{R} being the ratio of the reference speed $v_0$ to the phase speed $v_W$ in the medium. The same interpretation of $R$ has been used for the potential barrier case in Ref. \cite{galapon1}. 

We now focus on establishing the expected quantum traversal time $\tau_W$ across the potential well. We rewrite the complex index of refraction \textit{R}* by introducing the inverse Fourier transform $\varphi(q)=(1/2\pi)\int_{-\infty}^{\infty}\phi(\tilde{k})e^{i \tilde{k}q}d\tilde{k}$. Substituting $\varphi(q)$ into $\Phi(\zeta)$ given by Eq. (\ref{16th}) yields
\begin{equation}\label{19th}
\Phi(\zeta)=\int_{-\infty}^{\infty}d\tilde{k}\,|\phi(\tilde{k})|^2 e^{i \tilde{k}\zeta}.
\end{equation}
Substituting Eq. (\ref{19th}) into our definition of the complex index of refraction of the well $R^*$ and letting $\tilde{k}=k-k_0$ leads to
\begin{equation}\label{20th}
R^*=k_0\int_{0}^{\infty}d\zeta \, \int_{-\infty}^{\infty}dk |\phi(k-k_0)|^2 I_0(\kappa\zeta) e^{i k \zeta}.
\end{equation}
Notice that $\phi(k-k_0)$ is just the Fourier transform of the full incident wave function $\psi(q)=\varphi(q)e^{ik_0q}$, that is,
\begin{equation}
\phi(k-k_0)=\frac{1}{\sqrt{2 \pi}}\int_{-\infty}^{\infty}dq \,  e^{-ikq}\psi(q) = \tilde{\Psi}(k)
\end{equation}
The complex effective index of refraction of the well then takes the form 
\begin{equation}\label{21st}
R^*=k_0\int_{0}^{\infty}d\zeta \, I_0(\kappa\zeta)\int_{-\infty}^{\infty}dk \, |\tilde{\Psi}(k)|^2 e^{i k \zeta}.
\end{equation}

One may naively just interchange the order of integrations along $\zeta$ and $k$ in $R^*$, and then evaluate first the integral along $\zeta$. This leads to a divergent integral which can be normally evaluated by analytic continuation, regularization, summability methods, finite part integrals and others. However, we recently found out that naively interchanging the order of integrations in the real line leads us to miss some significant terms due to the special nature of divergent integrals \cite{tica,galapsmissed}. These missed terms naturally appear when one has to solve Eq. (\ref{21st}) in the complex plane. This problem is similar to but different from the problem of missing terms in term by term integration involving divergent integrals \cite{galapsmissed}. Now, evaluating Eq. (\ref{21st}) in the complex plane leads us to our final expression for the effective index of refraction of the well given by
\begin{equation}\label{25th}
\begin{aligned}
R=k_0&\int_{0}^{\infty} dk \,\frac{|\tilde{\Psi}(k)|^2}{\sqrt{k^2+\kappa^2}} - k_0\int_{0}^{\infty}dk \,\frac{|\tilde{\Psi}(-k)|^2}{\sqrt{k^2+\kappa^2}}\\
&-k_0 \, \int_{0}^{\kappa}dk \, \frac{\mbox{Im}\left[2\,\tilde{\Psi}(ik)\tilde{\Psi}^*(-ik)\right]}{\sqrt{\kappa^2-k^2}}.
\end{aligned} 
\end{equation}
The derivation of Eq. (\ref{25th}) is shown explicitly in Appendix \ref{app2}. Substituting Eq. (\ref{25th}) into $\tau_W=(L/v_0)R=(\mu L/\hbar k_0)R$ gives us the expected quantum traversal time across the potential well. 

Now, notice that the factor $\hbar\sqrt{k^2 +\kappa^2}/\mu$ can be defined as the velocity $v_{top}(k)$ on top of the potential well and $\hbar\sqrt{\kappa^2 - k^2}/\mu$ as the velocity $v_{in}(k)$ inside of the well. We can also define $\tau_{top}(k)=L/v_{top}(k)$ as the traversal times across the top of the well with velocity $v_{top}(k)$ and $\tau_{in}(k)=L/v_{in}(k)$ as the traversal times inside of the well with velocity $v_{in}(k)$ for a given \textit{k}. Hence, the expected quantum traversal time across the potential well is now given by
\begin{equation}\label{3.107}
\begin{split}
\tau_W = \int_{0}^{\infty}&dk \,\, \tau_{top}(k)|\tilde{\Psi}(k)|^2 - \int_{0}^{\infty}dk \,\, \tau_{top}(k)|\tilde{\Psi}(-k)|^2 \\
&- \, \int_{0}^{\kappa}dk \,\, \tau_{in}(k) \, \mbox{Im}\left[2 \,\tilde{\Psi}(ik)\tilde{\Psi}^*(-ik)\right].
\end{split}
\end{equation}

Equation (\ref{3.107}) is our final expression for the expected quantum traversal time across the potential well which is the main result of this paper. The first two terms of Eq. (\ref{3.107}) can be identified as the weighted sum of the classical traversal times $\tau_{top}(k)$ across the top of the potential well with weights $|\tilde{\Psi}(k)|^2$ and $|\tilde{\Psi}(-k)|^2$. In fact, these two weights can be identified as legitimate momentum probability distributions.  Furthermore, the first two terms of $R$ show the contribution from the positive and negative momentum components of the incident wave packet to the well traversal time. This is similar to but different from the result of Los et al. \cite{Los}. One may argue that only the positive components are relevant for arrivals at the transmission channel which is to be expected from the classical notion of traversal time. However, in quantum mechanics, a wave packet which represents an ensemble of particles cannot be initially localized in the region $q<0$ at $t=0$ if it only has positive momentum components. In fact, the negative momentum components of the incident wave packet, which by construction are restricted to the half-line in the momentum space, are not necessarily be equal to zero in the entire $q$ space $[-\infty,+\infty]$ and thus contribute \cite{Los}.  

Unlike the first two terms, the third term of Eq. (\ref{3.107}) has a pure quantum mechanical origin. Notice that the third term is expressed as a weighted sum of the traversal time inside the well with velocity $v_{in}(k)$ with weight $\mbox{Im}\left[2\, \tilde{\Psi}(ik)\tilde{\Psi}^*(-ik)\right]$. Unlike $|\tilde{\Psi}(k)|^2$ and $|\tilde{\Psi}(-k)|^2$, the weight $\mbox{Im}\left[2\tilde{\Psi}(ik)\tilde{\Psi}^*(-ik)\right]$ cannot be identified as a probability distribution because it is not positive definite and in some cases, has an oscillatory behavior. Recall that classically, the expectation is  that the incident particle will only traverse on top of the potential well because its incident energy $E$ is greater than the potential $-V_0$. Of course we know quantum mechanically that there is a portion of the incident wave packet that interacts with the potential well. This portion is captured by the well and remains trapped inside for a period of time. In other words, there is a nonzero probability that a quantum particle, described by a wave packet, is found inside of the potential. This phenomena is exactly what is depicted by the third term of Eq. (\ref{3.107}). The portion of the wave packet trapped inside the well traverses the well region and only comes out at time $\tau_{in} (k)$ for a given $k$. The contribution from this term becomes more apparent when one has to consider sufficiently large $\kappa$, that is, for deep potential wells. As will be shown later, the third term becomes dominant for deep wells implying a significant quantum mechanical interaction between the incident wave packet and the potential well. 

\section{Relation to the barrier traversal time}
Now that we have a closed-form expression of the quantum well traversal time, it would then be interesting to investigate how the latter is related to the known barrier traversal time found in Ref. \cite{galapon1}. We will show here that both the barrier and well traversal times vary smoothly when the potential amplitude is varied from $V_0 \to -V_0$ so that the two traversal times can be derived from each other. This also enables us to determine the origin of the zero tunneling time result of Ref. \cite{galapon1}.

\subsection{Well traversal time to barrier traversal time}\label{subsec:rwtorb}
\begin{figure*}
	\includegraphics[width=0.98\textwidth]{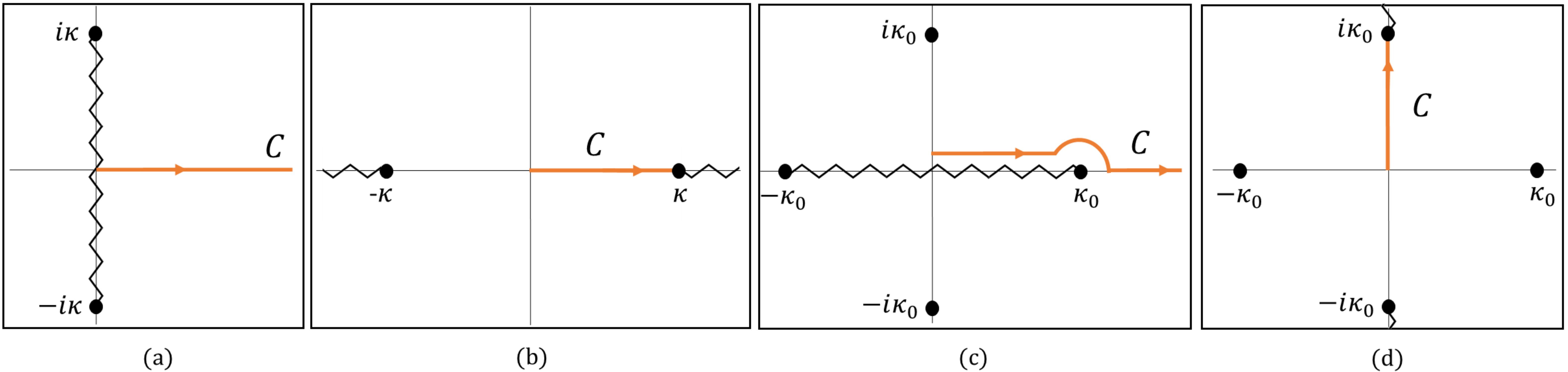}
	\centering
	\caption{The original contours of integration for the (a) first two terms and the (b) third term of the real-valued function $R$ given by Eq. (\ref{25th}). Here, $\kappa$ is a pure real number. The branch cuts of $1/\sqrt{k^2+\kappa^2}$ and $1/\sqrt{\kappa^2-k^2}$ are chosen such that they are analytic at $[0,\infty]$ and $[0,\kappa]$, respectively. When $\kappa$ becomes a pure imaginary number, $\kappa=i\kappa_0$, the chosen branch cuts are rotated counterclockwise. Hence, the previously real-valued function $R$ becomes now a complex-valued function $R(i\kappa_0)$ given by Eq. (\ref{riko}).  This implies new contour representations in the complex plane for the (c) first two terms and (d) third term of the refractive index $R(i\kappa_0)$.}
	\label{fig:rwtorb}
\end{figure*}

With the definition $\kappa=\sqrt{2\mu V_0}/\hbar$, the transformation $V_0 \to -V_0$ implies an equivalent transformation $\kappa \to i\kappa$. This then gives us the motivation to consider the previously real-valued function $R$ given by Eq. (\ref{25th}) as a complex-valued function so that the previously real-valued weights $|\tilde{\Psi}(\pm k)|^2$ and $\mbox{Im}\left[2\tilde{\Psi}(ik)\tilde{\Psi}^*(-ik)\right]$ are now to be considered entire functions of the complex variable $k$. 
We choose the branch cuts of $1/\sqrt{k^2+\kappa^2}$ and $1/\sqrt{\kappa^2-k^2}$ as depicted in Figs. (\ref{fig:rwtorb}.a) and (\ref{fig:rwtorb}.b), respectively. Now, we impose the transformation $V_0 \to -V_0$ by considering $\kappa$ to be a pure imaginary number, that is, $\kappa = i\kappa_0$ where $\kappa_0>0$. Equation (\ref{25th}) then leads to 
\begin{equation}\label{riko}
\begin{aligned}
R(i\kappa_0)=&k_0\int_{0}^{\infty} dk \,\frac{|\tilde{\Psi}(k)|^2}{\sqrt{k^2-\kappa_0^2}} - k_0\int_{0}^{\infty}dk \,\frac{|\tilde{\Psi}(-k)|^2}{\sqrt{k^2-\kappa_0^2}}\\
&-k_0 \, \int_{0}^{i\kappa_0}dk \, \frac{\mbox{Im}\left[2\,\tilde{\Psi}(ik)\tilde{\Psi}^*(-ik)\right]}{\sqrt{k^2+\kappa_0^2}}
\end{aligned} 
\end{equation}
where the corresponding contour representations in the complex plane are shown in Figs. (\ref{fig:rwtorb}.c) and (\ref{fig:rwtorb}.d). Note that $\tilde{\Psi}(ik)\tilde{\Psi}^*(-ik)=|\tilde{ \Psi}(\pm z)|^2|_{z=\pm ik}$ so that the factor $\mbox{Im}\left[2\,\tilde{\Psi}(ik)\tilde{\Psi}^*(-ik)\right]$ in the third term of Eq. (\ref{riko}) can be expressed as a sum involving the real and imaginary parts of $|\tilde{ \Psi}(\pm z)|^2|_{z=\pm ik}$. Using the contour shown in Fig (\ref{fig:rwtorb}), one finds that the third term of $R(i\kappa_0)$ can be rewritten as
\begin{equation}\label{riko3rd}
R_{3}(i\kappa_0)=-k_0\int_{0}^{\kappa_0} dk \,\frac{|\tilde{\Psi}(k)|^2}{\sqrt{k^2-\kappa_0^2}} + k_0\int_{0}^{\kappa_0}dk \,\frac{|\tilde{\Psi}(-k)|^2}{\sqrt{k^2-\kappa_0^2}}
\end{equation}
which exactly cancels the $[0,\kappa_0]$ limit of the first two integrals of Eq. (\ref{riko}). This then leads to $R(i\kappa_0)=R_B$ where
\begin{equation}\label{rikobarrier}
R_B=-k_0\int_{\kappa_0}^{\infty} dk \,\frac{|\tilde{\Psi}(k)|^2}{\sqrt{k^2-\kappa_0^2}} + k_0\int_{\kappa_0}^{\infty}dk \,\frac{|\tilde{\Psi}(-k)|^2}{\sqrt{k^2-\kappa_0^2}}
\end{equation}

With $\kappa_0=\kappa=\sqrt{2 \mu V_0}/\hbar$, one can identify $R_B$ as the same barrier effective index of refraction found in Ref. \cite{galapon1}. Equation (\ref{rikobarrier}) then implies that only those components of $\tilde{\Psi}(\pm k)$ with $|k|>\kappa$ contribute to any measurable traversal time across the barrier region. For the case of quantum tunneling, the support of the momentum distribution of the incident wave packet has a corresponding energy distribution that lies below the potential height. This then implies that the index of refraction is zero and the traversal time under the barrier vanishes. This inevitably leads to the conclusion
that below the barrier energy components are transmitted
without delay across the barrier, that is, quantum tunneling happens instantaneously \cite{galapon1}. 

Our result here suggests that the well index of refraction leads exactly to the barrier refractive index when the potential value $V_0$ is varied to $-V_0$. Since $\tau_B=(L/v_0)R_B$, this immediately implies that the well traversal time also leads to the barrier traversal time under the same variation. This is interesting because the classical well traversal time cannot be generally extended to the barrier traversal time for all energy values of the incident particle because the latter is undefined when $E_0<V_0$. This stems from the fact that quantum tunneling violates the principles of classical mechanics. However, we have shown here that the quantum well traversal time varies smoothly such that the barrier traversal time can be derived from the former under the transformation $V_0 \to -V_0$. This, in fact, origins from the wave-like aspect of the incident particle, and the prediction of quantum tunneling in quantum mechanics. 

In addition, what we fail to see in Ref. \cite{galapon1} which we have seen here is that the barrier traversal time can actually be written as the sum involving the classical and quantum contributions, similar to the well traversal time given by Eq. (\ref{3.107}) as discussed in Sec. \ref{sec:level3}. However, the presumably third term in $R_B$ with weight $\mbox{Im}\left[2\tilde{\Psi}(ik)\tilde{\Psi}^*(-ik)\right]$, which we identified earlier as a purely quantum contribution to the traversal time, exactly cancels the nonclassical contribution of the supposedly finite quantum tunneling time (that is, when $\tilde{\Psi}(\pm k)$ with $|k|>\kappa$). Hence, we only have above-the-barrier traversal time. This cancellation may physically originate from the interaction of the wavepacket components, such as interference and multiple reflections, happening inside the potential barrier. 

\subsection{Barrier traversal time to well traversal time}\label{subsec:rbtorw}
We have shown in the previous subsection that the well traversal time leads correctly to the barrier traversal time when $V_0$ varies to $-V_0$. For completeness, we also show that the barrier traversal time also leads correctly to the well traversal time under the same transformation. 

\begin{figure*}
	\includegraphics[width=0.98\textwidth]{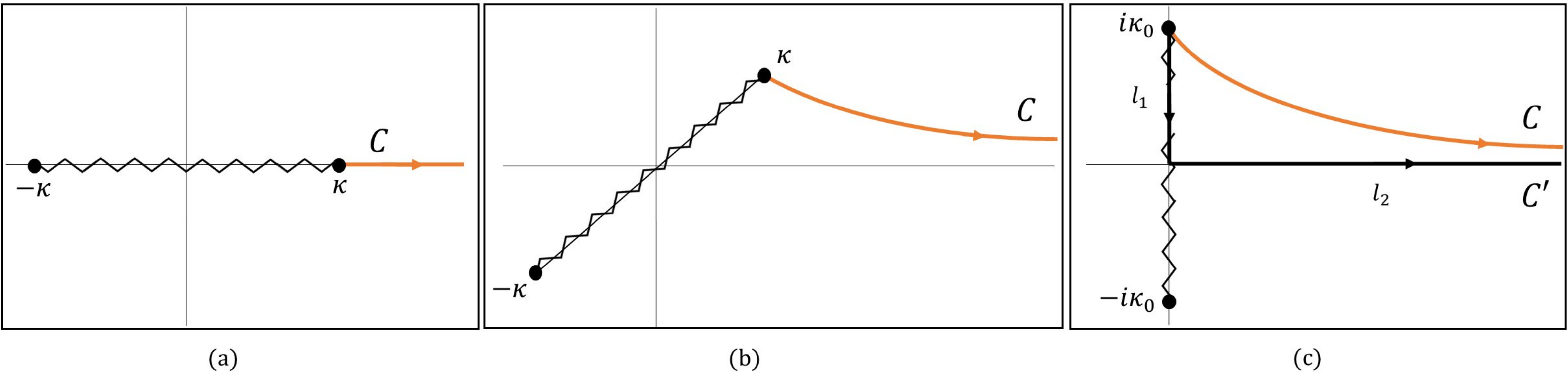}
	\centering
	\caption{The contour of integration $C$ for $R_B$ when (a) $\kappa$ is a purely real number. (b) $\kappa$ is generally a complex number, and (c) $\kappa$ is a purely imaginary number given by $\kappa=i\kappa_0$. The contour $C$ is then deformed to an equivalent contour $C'$ with line segments $l_1$ and $l_2$. }
	\label{fig:rbtorw}
\end{figure*}

We impose similar assumptions used in the previous subsection on $R_B$ given by Eq. \ref{rikobarrier}, $|\tilde{\Psi}(\pm k)|^2$ and $\mbox{Im}\left[\tilde{\Psi}(ik)\tilde{\Psi}^*(-ik)\right]$ and lift the integrals into the complex plane. For $k>0$, the branch of $1/\sqrt{k^2 - \kappa^2}$ can be chosen such that it is analytic in the interval $[\kappa,\infty]$ as shown in Fig. (\ref{fig:rbtorw}.a). The branch cut is set between the branch points $\pm \kappa$. When $k$ is a general complex number, we can choose the contour $C$ shown in Fig. (\ref{fig:rbtorw}.b) as our contour of integration so that it does not go around the branch cut. Now, if $\kappa=i\kappa_0$, Eq. (\ref{rikobarrier}) leads to 
\begin{equation}\label{rbarrier3}
R_B=k_0\int_{i\kappa_0}^{\infty} dk \,\frac{|\tilde{\Psi}(k)|^2}{\sqrt{k^2+\kappa_0^2}} - k_0\int_{i\kappa_0}^{\infty}dk \,\frac{|\tilde{\Psi}(-k)|^2}{\sqrt{k^2+\kappa_0^2}}
\end{equation}
where $\kappa_0>0$ and the corresponding contour of integration is the contour $C$ in Fig. (\ref{fig:rbtorw}.c). However, Eq. (\ref{rbarrier3}) can also be rewritten by deforming the contour $C$ into an equivalent contour $C'$ so that the two integrals are evaluated along the segments $l_1$ and $l_2$ shown in Fig. (3b). One then finds
\begin{equation}\label{repsilon2}
\begin{split}
\int_{i\kappa_0}^{\infty} dk \frac{| \tilde{\Psi}( k)|^2}{\sqrt{k^2+\kappa_0^2}}&=\int_{0}^{\infty} dk \frac{|\tilde{\Psi}( k)|^2}{\sqrt{k^2+\kappa_0^2}}\\
&- i\int_{0}^{\kappa_0}dk \frac{\tilde{\Psi}(ik)\tilde{\Psi}^*(-ik)}{\sqrt{\kappa_0^2-k^2}}\\
\int_{i\kappa_0}^{\infty} dk \frac{| \tilde{\Psi}(- k)|^2}{\sqrt{k^2+\kappa_0^2}}&=\int_{0}^{\infty} dk \frac{|\tilde{\Psi}(- k)|^2}{\sqrt{k^2+\kappa_0^2}}\\
&+ i\int_{0}^{\kappa_0}dk \frac{\left[\tilde{\Psi}(ik)\tilde{\Psi}^*(-ik)\right]^*}{\sqrt{\kappa_0^2-k^2}}.
\end{split}
\end{equation}

From Eqs. (\ref{rbarrier3}) and 	(\ref{repsilon2}), we see that the barrier refractive index also leads exactly to the well refractive index given by Eq. (\ref{25th}). This immediately implies that the barrier traversal leads to the well traversal time as $V_0 \to -V_0$. We can then conclude that both the barrier and well traversal times varies smoothly such that the two can be derived from each other when $V_0 \to -V_0$.

\section{\label{sec:level4} well traversal Time for Gaussian wave packets}
To better appreciate Eqs. (\ref{25th}) and (\ref{3.107}), we consider an incident Gaussian wave packet of the form 
\begin{equation}\label{30th}
\varphi(q)=\frac{1}{\sqrt{\sigma\sqrt{2\pi}}}e^{-(q-q_0)^2/4\sigma^2},
\end{equation} 
where $q_0$ and $\sigma^2$ are the initial position and position variance, respectively. In momentum space representation, $\varphi(q)$ is expressed as
\begin{equation}\label{29th}
\tilde{\Psi}(k)=\sqrt{\sigma\sqrt{\frac{2}{\pi}}}e^{-iq_0(k-k_0)}e^{-\sigma^2(k-k_0)^2}. 
\end{equation}
Hence, we have the following quantities in Eq. (\ref{25th}),
\begin{equation}\label{psi2}
|\tilde{\Psi}(\pm k)|^2=\sqrt{\frac{2}{\pi}}\sigma e^{-2\sigma^2(k \mp k\textsubscript{0})^2}
\end{equation}
\begin{equation}\label{psii}
\tilde{\Psi}(ik)\tilde{\Psi}^*(-ik)=\sqrt{\frac{2}{\pi}}\sigma e^{-2\sigma^2(ik-k\textsubscript{0})^2}.
\end{equation}
Using Eqs. (\ref{psi2}) and  (\ref{psii}), the effective index of refraction of the well can be written in the form
\begin{equation}
R=R_++R_-+R_\kappa
\end{equation}
where 
\begin{equation}\label{}
R_{\pm} = \pm k\textsubscript{0}\sigma\sqrt{\frac{2}{\pi}}\int_{0}^{\infty} dk \,\frac{e^{-2\sigma^2(k \mp k_0)^2}}{\sqrt{k^2+\kappa^2}}
\end{equation}
\begin{equation}\label{r3rd}
R_{\kappa}=-2 k_0 \sigma \sqrt{\frac{2}{\pi}} \mbox{Im} \int_{0}^{\kappa} dk \, \frac{e^{2 \sigma^2 (k+ik_0)^2}}{\sqrt{\kappa^2-k^2}}    
\end{equation}
The indices $+$ and $-$ indicate the contributions from the positive and negative momentum components of the incident wave packet, respectively. The third term with index $\kappa$ describes the significant contribution of the well depth to $R$. Equation (\ref{r3rd}) is rewritten by lifting the integral in the complex plane. This leads to 
\begin{equation}\label{rkapppa}
\begin{split}
R_{\kappa}=&-2 k_0 \sigma \sqrt{\frac{2}{\pi}}\, \bigg[e^{2 \sigma^2 (\kappa^2-k_0^2)}\, \mbox{Im} (z)+ e^{-2 \sigma^2 k_0^2}\gamma \bigg]
\end{split}
\end{equation}
where
\begin{equation}\label{eqz}
z= e^{4i\sigma^2 k_0 \kappa} \,\int_{0}^{\infty} i \, dk \, \frac{e^{-2 \sigma^2 (k^2 + 2 k_0 k)}e^{4 i \sigma^2 \kappa k}}{\sqrt{k^2-2i \kappa k}},
\end{equation}
\begin{equation}
\gamma=\int_{0}^{\infty}  \, dk \, \frac{e^{-2 \sigma^2 (k^2 - 2 k_0 k)}}{\sqrt{k^2+\kappa ^2}}.
\end{equation}
One immediately notices from Eq. (\ref{r3rd}) that the third term of $R$ approaches $0$ as $\kappa \to 0$. However, this becomes dominant when $\kappa \to \infty$, $\sigma \to \infty$ when $\kappa>k_0$, or both as can be seen from Eq. (\ref{rkapppa}) because of the exponentially large factor $e^{2 \sigma^2 (\kappa^2-k_0^2)}$. In fact, $|R_\kappa|$ becomes sufficiently large enough so that the contibutions from the first two terms of $R$ become negligible. That is, $R \sim R_\kappa$  in the limit $\kappa \to \infty$. Furthermore, the sign of $R_\kappa$ oscillates from positive to negative for specific values of $\sigma k_0$ and $\sigma \kappa$ which will be shown later. This then implies that the Gaussian wave packet, on average, can be either advanced or delayed as it is transmitted through and reflected by the potential well. 

\subsection{Shallow potential wells} 
For this paper, shallow potential wells are described using the condition $\kappa/k_0 \to 0$. Two distinct and interesting results are found when one considers spatially wide and narrow Gaussian wave packets. 	
All results are numerically verified using random values of $\kappa$, $k_0$, and $\sigma$. 

\subsection*{Spatially wide incident Gaussian wave packets} 
\begin{figure}
	\includegraphics[width=.48\textwidth]{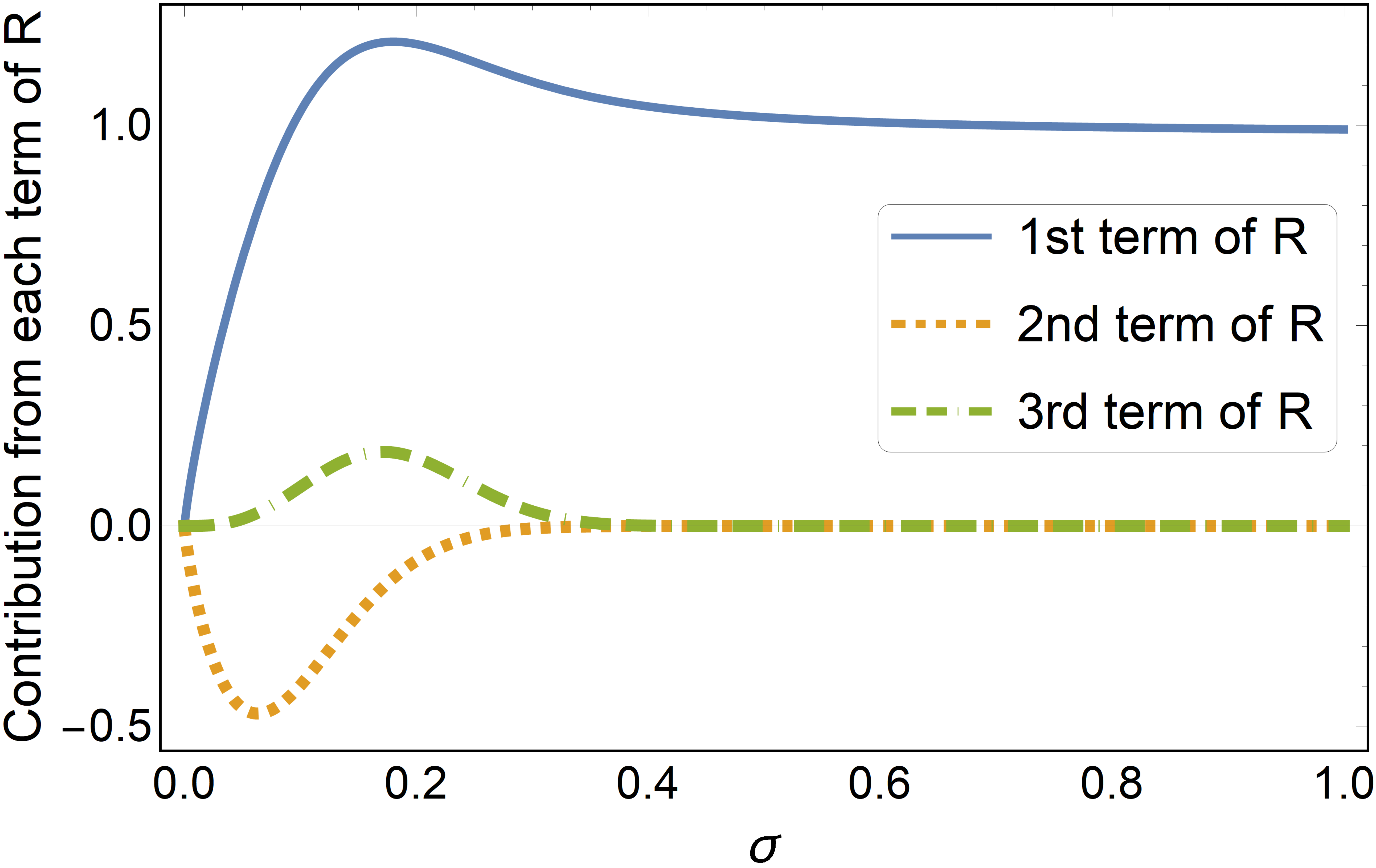}
	\centering
	\caption{The contribution from each term of $R$ as a function of $\sigma$ where $k_0=5$ a.u. and  $\kappa=1$ a.u.. The effective index of refraction $R$ of the well approaches a constant value as $\sigma$ increases. This constant value is exactly the classical refractive index of the well. Hence, the quantum well traversal time approaches the classical well traversal time.}
	\label{fig:sigma}
\end{figure}

When $\sigma \to \infty$, our momentum distribution is localized. The contributions from the second and third terms in Eq. (\ref{25th}) are negligible compared to the first term in the first approximation since the factor $e^{-2\sigma^2k_0^2} \to 0$ much faster than the integrals along $k$. This behavior can be easily seen in the plot for the refractive index $R$ as a function of the spread $\sigma$ shown in Fig. (\ref{fig:sigma}). The 1st term of $R$ dominates when $\sigma$ is large and the last two terms are negligible. This suggests that the contribution from the positive momentum component is only relevant to the traversal time in the transmission channel. This physically signifies that in the limit $\sigma \to \infty$, the wave packet becomes a pure plane wave such that the traversal time is only due to the positive momentum component of the incident wave packet. In fact, this is what happens classically. 

Now, since our momentum distribution is sharply peaked around its central point, the integral for the first term of $R$ can be extended from $[0,+\infty]$ to $[-\infty,+\infty]$. Substituting $\tilde{\Psi}(k)$ to the first term of Eq. (\ref{25th}), expanding the factor $1/\sqrt{k^2+\kappa^2}$ in Taylor series about $k=k_0$, interchanging the order of summation and integration, and integrating term by term, we find 
\begin{equation}\label{32nd}
R \sim \frac{k_0}{\sqrt{k_0^2 + \kappa^2}} + \sum_{n=1}^{\infty}\frac{1}{(2\sigma)^{2n}}\frac{(2n-1)!!}{(2n)!} \chi_n
\end{equation}
where
\begin{align}
	\chi_n=\frac{d^{(2n)}}{dk^{(2n)}}(k^2 + \kappa^2)^{-\frac{1}{2}}\bigg|_{\textsubscript{$k=k_0$}}
\end{align}
Substituting $k_0=\sqrt{2\mu E_0}/\hbar$ and $\kappa=\sqrt{2\mu V_0}/\hbar$ , the first term of Eq. (\ref{32nd}) is equal to $\sqrt{E_0/(E_0+V_0)}$, which is exactly Eq. (\ref{18th}), the effective refractive index of the potential well in the high enegy limit. Substituting $R$ in $\tau_W$ leads to the quantum traversal time across the well given by
\begin{equation}\label{33rd}
\tau_W \sim \frac{\mu L}{\sqrt{p_0^2+2\mu V_0}}+\bigg(\frac{\mu L}{\sqrt{2\mu E _0}}\bigg)\sum_{n=1}^{\infty}\frac{1}{(2\sigma)^{2n}}\frac{(2n-1)!!}{(2n)!} \chi_n
\end{equation}

Notice that the first term of Eq. (\ref{33rd}) is exactly equal to Eq. (\ref{eigth}) which is the classical traversal time across the potential well. The succeeding terms are the quantum corrections to the classical traversal time. These quantum corrections are dependent to $\hbar$, $\sigma$ and $\kappa$. As $\hbar \to 0$ or in the extreme limit $\sigma \to \infty$, the quantum corrections vanish and the expected quantum traversal time  becomes exactly the classical traversal time, which depicts the known correspondence principle. This result is also seen in Fig. (\ref{fig:sigma}). The plot for the first term of $R$ approaches a constant value as $\sigma$ increases and this constant value is exactly the classical index of refraction of the well. 

An important implication of this result is that the size of the incident wave packet determines the nonclassicality of the expected traversal time across the well over mass and incident energy. This is a manifestation of the quantum wave packet size effect which is also found by one of us for the free particle case \cite{galaponsize}. This then serves as an extension of the said quantum mechanical effect from the free particle case to the potential well case. We can then use a similar discussion made by Galapon for the free case \cite{galaponsize}. One can say that the more a particle is quantum in nature, the more prominent is wave-property is. The more classical is its traversal time, the more the particle is classical in nature. The more the particle is localized, the more nonclassical is its traversal time \cite{galaponsize}.

For this particular case where the arrival at the transmission channel is dominated by the contribution from the positive momentum component of the incident wave packet, the measurable quantum traversal time is given by 
\begin{equation}\label{4.23}
\tau_W \sim \int_{-\infty}^{\infty}dk \, \tau_{top}(k)|\tilde{\Psi}(k)|^2
\end{equation}
which implies that $\tau_W$ is just the weighted sum of the classical traversal times $\tau_{top}(k)$ on top of the well with weights $|\tilde{\Psi}(k)|^2$.

\subsection*{Spatially narrow incident Gaussian wave packets}
\begin{figure}
	\includegraphics[width=.48\textwidth]{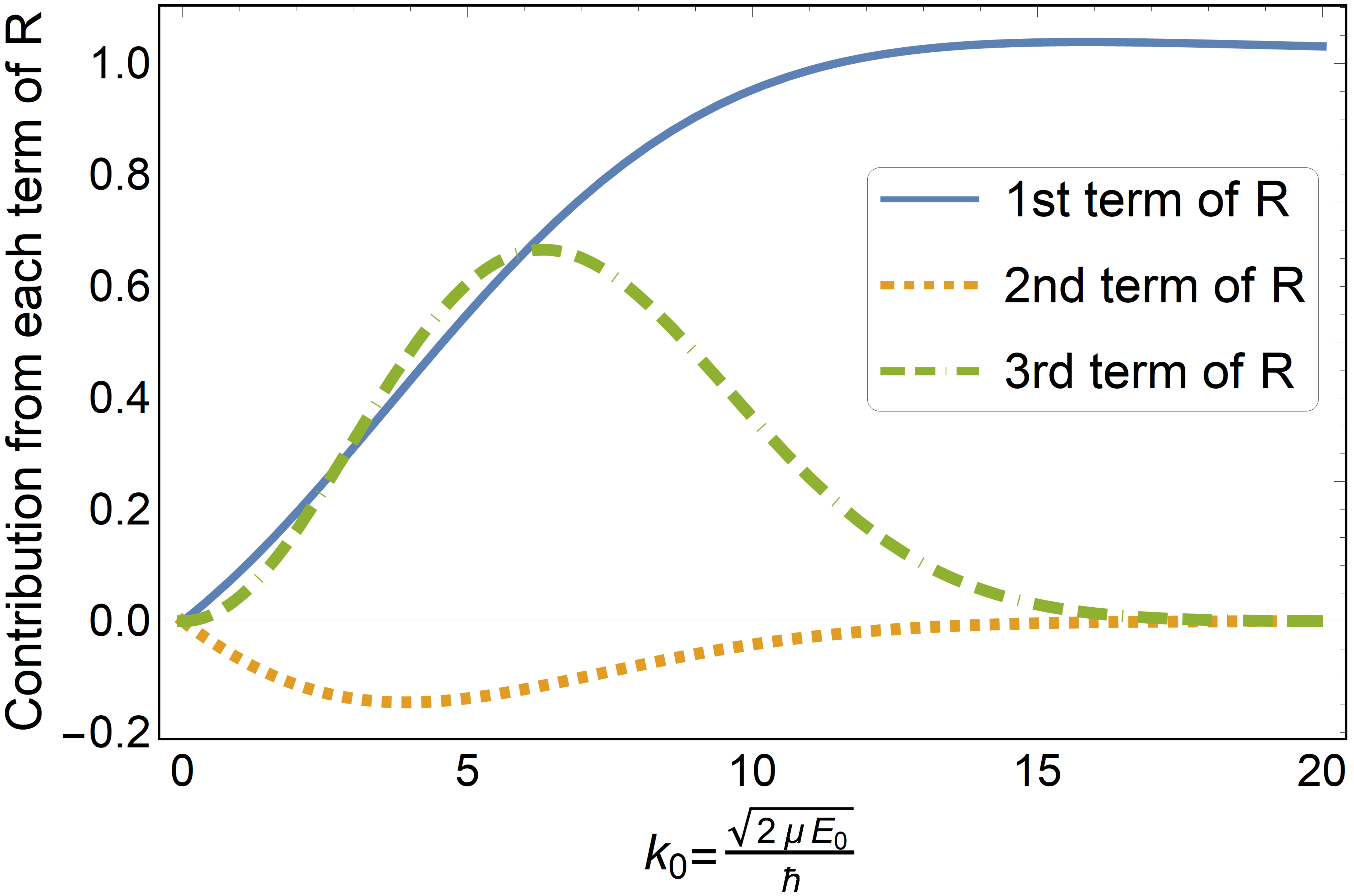}
	\centering
	\caption{The contribution from each term of $R$ as a function of $k_0$ where $\sigma=1/10$ a.u. and $\kappa=5$ a.u..}
	\label{fig:k0varies}
\end{figure}

For this case, we are to consider $\sigma \to 0$, and $\kappa/k_0 \to 0$. Since $\sigma$ is very small, our momentum distribution is broad and we can no longer neglect the contributions from the second and third terms in Eq. (\ref{25th}) as we did previously. This can be verified if we plot $R$ as shown in Fig. (\ref{fig:k0varies}). The third term of Eq. (\ref{25th}) significantly contributes to the effective index of refraction of the well when $\kappa/k_0 \le 1$ but becomes subdominant when $\kappa/k_0 \to \infty$. This is because the third term is only dependent to the components $0 \le k \le \kappa$ and approaches to zero when $\kappa \to 0$. 
Using Eqs. (\ref{psi2}) and (\ref{psii}), Eq. (\ref{25th}) leads to \begin{equation}\label{36th}
\begin{split}
R=&2\sqrt{\frac{2}{\pi}}k_0 \sigma e^{-2\sigma^2k_0^2}\int_{0}^{\infty} dk\, \frac{e^{-2\sigma^2k^2}\mbox{sinh}(4 \sigma^2 k_0 k)}{\sqrt{k^2+\kappa^2}}\\
&+ 2\sqrt{\frac{2}{\pi}}k_0 \sigma e^{-2\sigma^2k_0^2} \,\, \mbox{Im} \int_{0}^{\kappa} dk\, \frac{e^{2\sigma^2(k+ik_0)^2}}{\sqrt{k^2+\kappa^2}}.
\end{split}
\end{equation}
We expand the hyperbolic sine and exponential functions in Eq. (\ref{36th}) in Taylor series, interchange the order of integration and summation and then perform a term-by-term integration which leads to an infinite series involving hypergeometric functions. 

However, recall that $\sigma$ and $\kappa$ are small, so we can expand these hypergeometric functions about $\sigma\kappa=0$, and retain the dominant terms. This leads to $R \sim Q+\Omega$ where
\begin{equation}\label{41st}
\Omega \sim \sqrt{2\pi}k_0 \sigma e^{-2 k_0^2 \sigma^2} \bigg[k_0 \textbf{L}_0(\alpha)-k_0 \bm{\mathcal{H}}_0(\alpha)+\kappa \textbf{L}_1(\alpha)\bigg]
\end{equation}
\begin{equation}\label{42nd}
Q=\sqrt{2\pi}k_0 \sigma e^{-2 k_0^2 \sigma^2}\mbox{erfi}  (\sqrt{2 k_0 \sigma}),
\end{equation}
$\alpha=4k_0\kappa \sigma^2$, $\bm{\mathcal{H}}_\nu(z)$ is the Struve function and $\textbf{L}_\nu(z)$ is the modified Struve function. Notice that Eq. (\ref{42nd}) is exactly the quantum correction factor to the classical expected time of arrival for a free particle found in \cite{galaponsize}. Since the expected quantum traversal time is $\tau_W=(L/v_0)R$, we find \begin{equation}\label{43rd}
\tau_W \sim Q\frac{\mu L}{\hbar k_0} + \Omega \frac{\mu L}{\hbar k_0}
\end{equation}
for sufficiently small $\sigma$ and $\kappa$. The first term of Eq. (\ref{43rd}) is exactly the expected quantum traversal time for the free particle found by one of us in Ref. \cite{galaponsize} for the same incident Gaussian wave packet. The second term of Eq. (\ref{43rd}) gives the quantum corrections to the expected quantum traversal time for the free particle in the well region. These corrections are all dependent on $\kappa$, i.e., the depth of the potential well. As we decrease $\kappa$, the contribution from these terms decreases and in the extreme limit $\kappa \to 0$, Eq. (\ref{43rd}) becomes exactly the quantum free traversal time found in \cite{galaponsize}. This result is to be expected if the concept of traversal time is extended to the implications of perturbation theory in quantum mechanics. A shallow potential well can be viewed as a weak perturbation to our free system and induces corrections to the quantum traversal time for a free particle. The same result is found if one is to solve Eq. (\ref{25th}) by using the newly proposed finite-part integration of the generalized Stieltjes transform found in \cite{tica}.	 
\subsection{Deep potential wells}

\begin{figure}
	\includegraphics[width=0.45\textwidth]{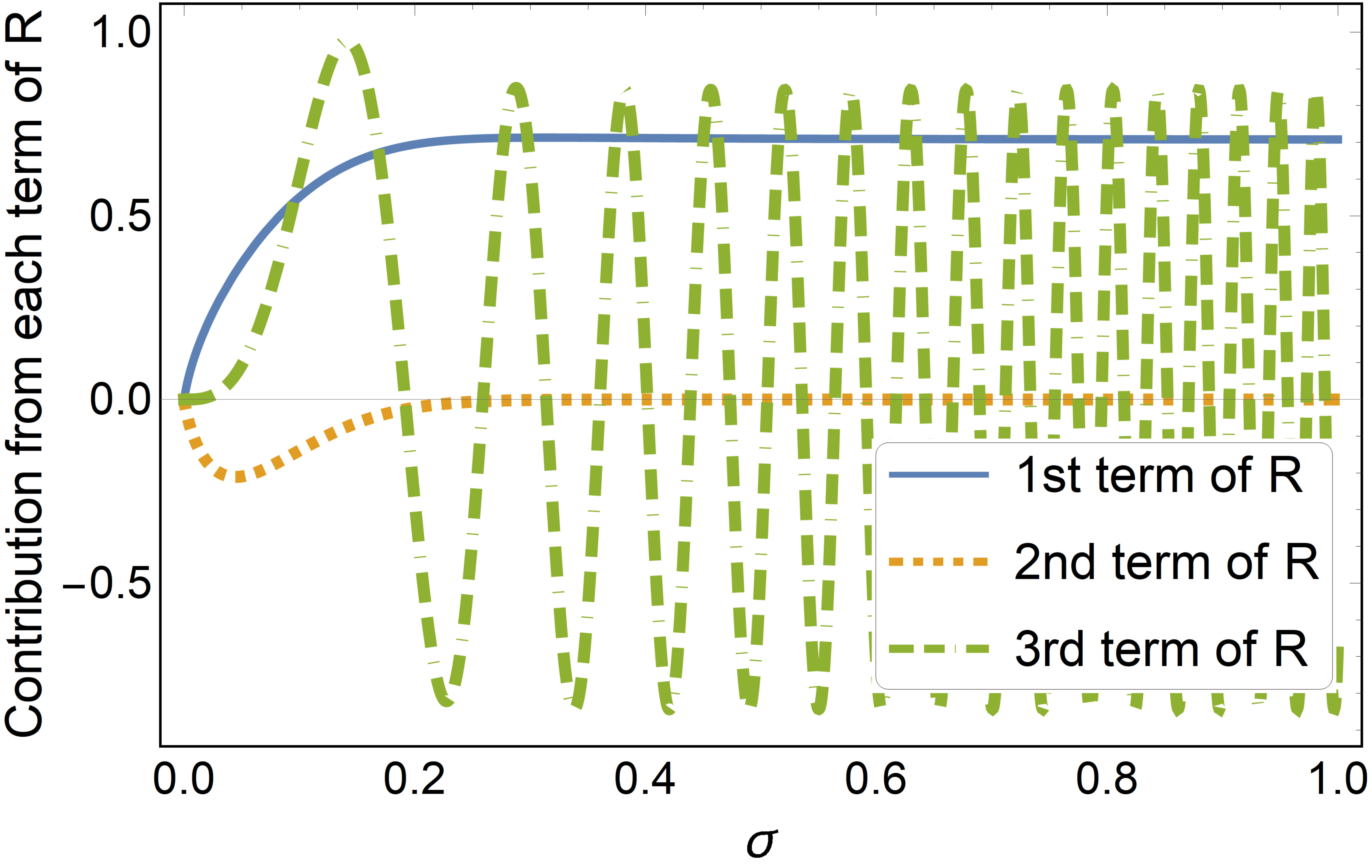}
	\centering
	\caption{The contribution from each term of $R$ as a function of $\sigma$ where $k_0=\kappa=5$ a.u. The first term of R still approaches the classical index of refraction as $\sigma$ increases. The third term of R now oscillates from positive to negative.}
	\label{fig:same}
\end{figure}

As we move away from the high energy limit $k_0 \to \infty$, the contribution from the third term of $R$ increases. And in the case when the incident energy $E_0$ becomes equal to the the potential depth $V_0$, i.e. $k_0=\kappa$, the magnitude of the third term now becomes greater than the first term. In fact, the complete index of refraction $R$ starts to oscillate from positive to negative and its oscillation becomes more rapid as $\sigma$ increases. 
Now for sufficiently deep wells, $\kappa/k_0$ becomes arbitrarily large. Since $\kappa$ is very large, the first and second terms of $R$ are negligible while the third term of $R$ is exponentially large because of the factor $e^{2\sigma^2 \kappa^2}$. From Eq. (\ref{rkapppa}), the refractive index of refraction for deep wells is 
\begin{equation}\label{rlarge}
R_\kappa \sim -2 \sqrt{\frac{2}{\pi}} k_0 \sigma \, e^{2 \sigma^2 (\kappa^2-k_0^2)}  \, \mbox{Im}(z),
\end{equation}
where $z$ is given by Eq. (\ref{eqz}).

\begin{figure}
	\includegraphics[width=0.47\textwidth]{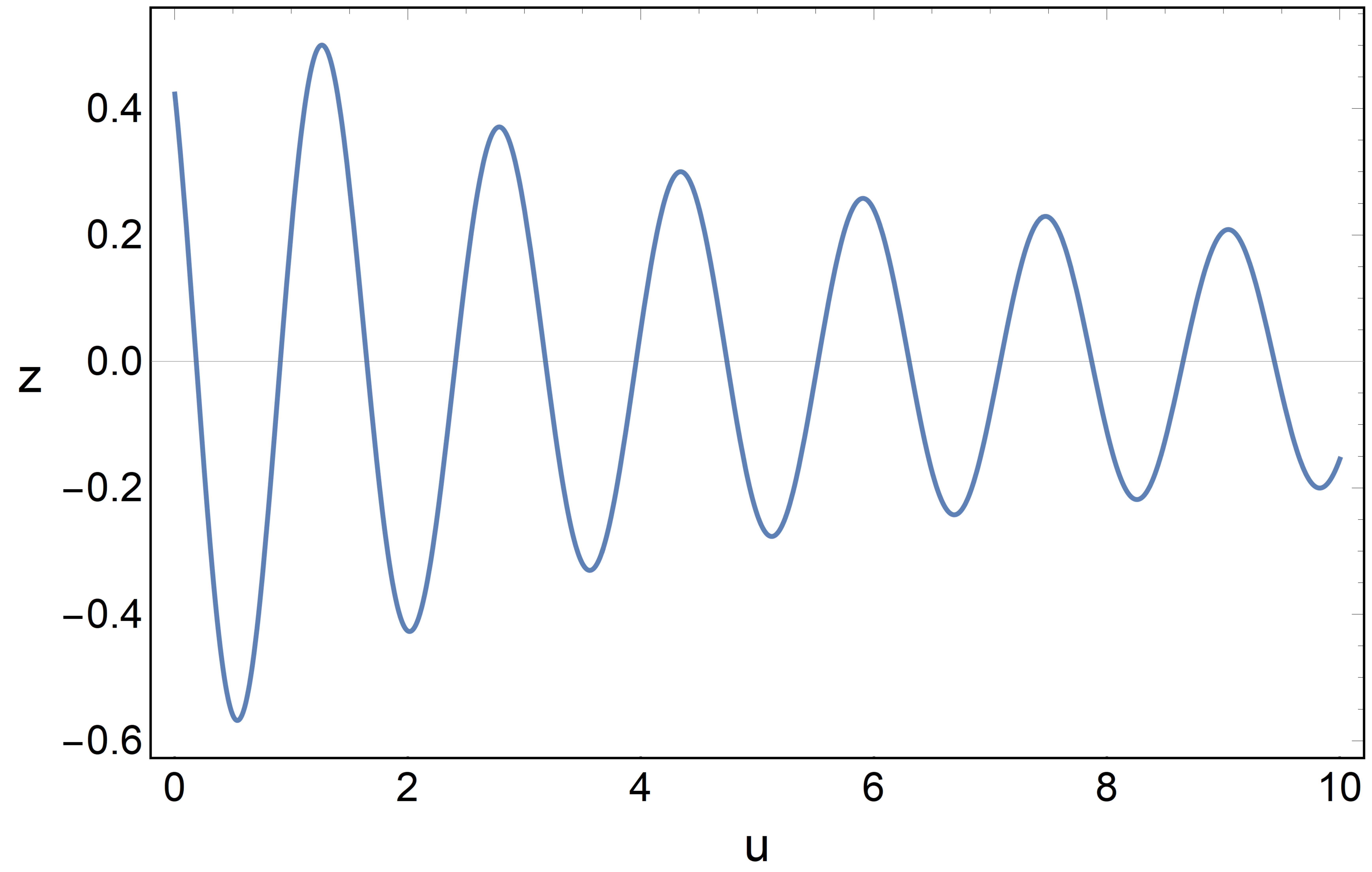}
	\centering
	\caption{Two-dimensional plot for the imaginary part of the factor $z$ where $v=1$. This oscillatory behavior suggests that the effective index of refraction $R$ of the potential well can be positive or negative. }
	\label{fig:v_1}
\end{figure}

\begin{figure}
	\includegraphics[width=0.45\textwidth]{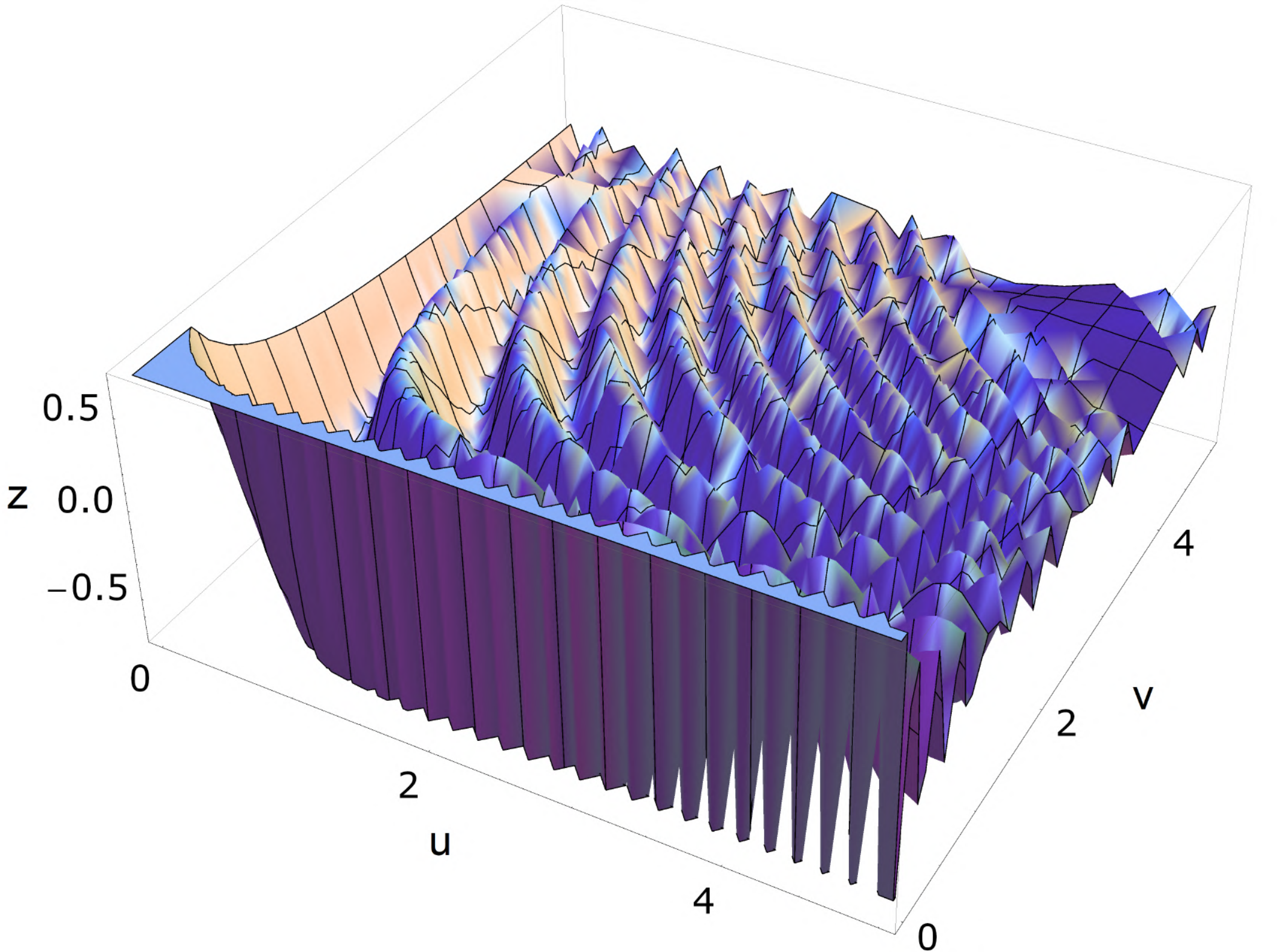}
	\centering
	\caption{Three-dimensional plot of the imaginary part of $z$. The same oscillatory behavior is observed implying a positive or negative index of refraction $R$.}
	\label{fig:3d}
\end{figure}

For convenience, we change variables to $u=\sigma \kappa$ and $v=\sigma k_0$ and numerically solve $z$. One can already notice that $R$ oscillates from positive to negative for deep potential wells because of the presence of the complex exponential $e^{4 i \sigma^2 k_0 \kappa}$ in Eq. (\ref{eqz}). This can be further seen from the $2d$ and $3d$ plots of the imaginary part of $z$ shown in Figs. (\ref{fig:v_1}) and (\ref{fig:3d}), respectively. 
\begin{figure}
	\includegraphics[width=0.45\textwidth]{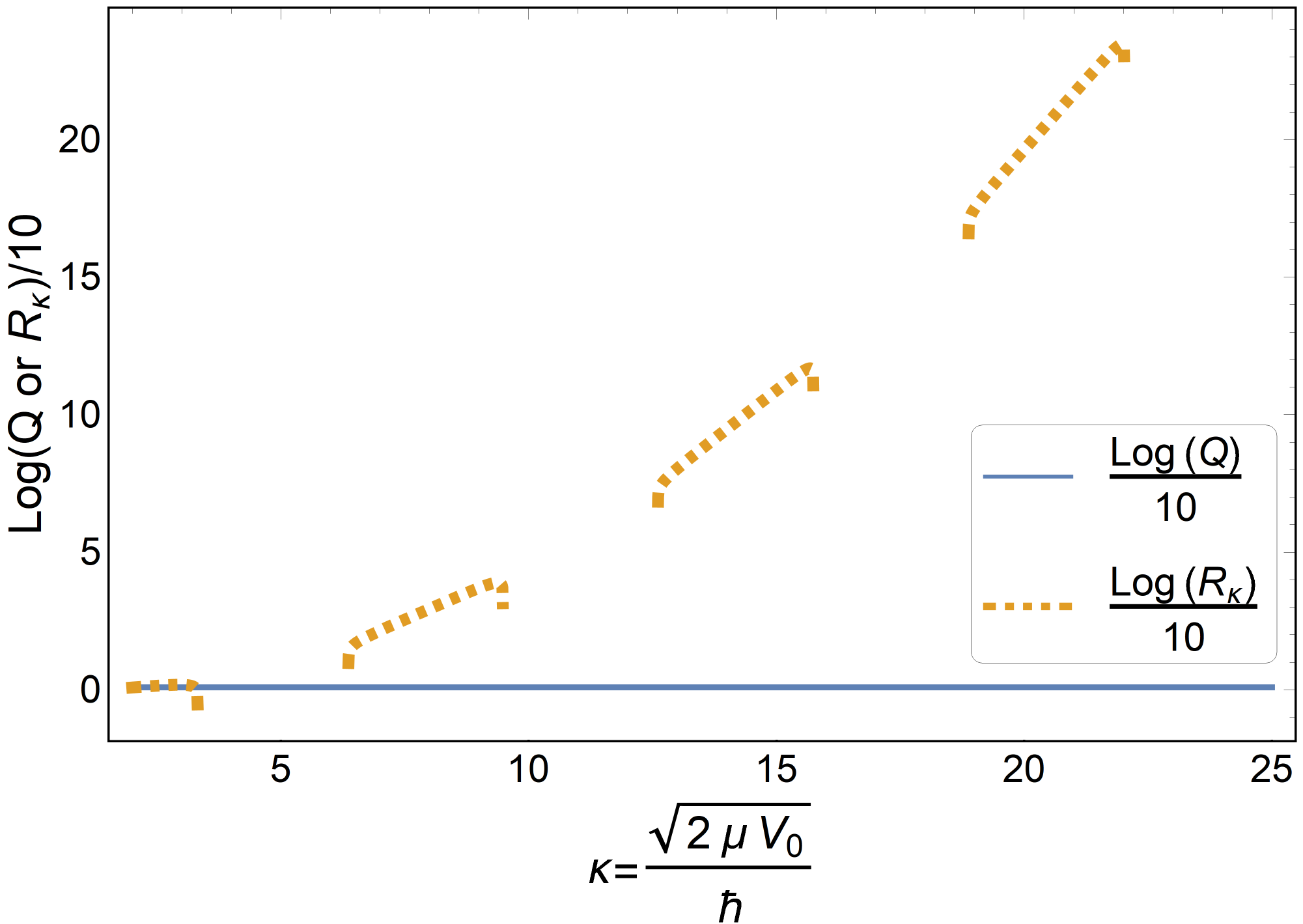}
	\centering
	\caption{Comparison between the correction factor $Q$ for the free particle and the index of refraction $R$ of the well in log scale. The discontinuity in the plot of $\mbox{log}(R_\kappa)/10$ happens when $R_\kappa$ is negative which implies that the corresponding traversal time is negative.}
	\label{fig:compare}
\end{figure}

Figures (\ref{fig:v_1}) and (\ref{fig:3d}) imply that the TOA difference $\Delta \tau=\tau_F-\tau_W$ can be positive or negative bacause of the oscillatory behavior of  $\tau_W=(L/v_0)R$ for deep potential wells. This can be further seen in Fig. (\ref{fig:compare}). The plot for the correction factor $Q$ for the free particle is negligible compared to the plot for the refractive index $R_\kappa$ for deep wells in logarithmic scale, that is, $|R_\kappa|>Q$ for $\kappa \to \infty$. 

Furthermore, notice that the plot for  $\mbox{log}(R_\kappa)/10$ is discontinuous and the discontinuities arise because $R_\kappa$ is negative. These discontinuities show the values of $\kappa$ for fixed $\sigma$ and $k_0$ wherein the quantum particle travels faster in the well region (advanced) compared to a free region. 

One interpretation to the negative index of refraction found here is that the quantum particle or wave packet is already at the arrival point prior to time $t = 0$ and is now moving away from it. However, this interpretation seems to be physically unacceptable if this is to be taken literally. A more acceptable interpretation is that the wave packet propagates with negative group velocity upon hitting the potential well. This implies that the wave packet is being reflected away and moving to the left from the potential well system. Hence, the wave packet is not detected at the arrival point. This is similar but different from the nonevanescent propagation with negative phase shifts prediction of Li and Wang for particles passing through a potential well \cite{Li}. Whatever the negative traversal time means for the well case, it will be meaningless to compare it to the free arrival time, which, for our setup, is always positive. 

On the other hand, the visible plot for $R_\kappa$ show the values of $\kappa$ for fixed $\sigma$ and $k_0$ wherein the particle travels slower in the well region (delayed). In principle, one can determine the specific values of  $\sigma$, $k_0$, and $\kappa$ when the particle is delayed ($R_\kappa>Q$) and advanced ($R_\kappa<Q$) for deep potential wells. This result may have an interesting application to microelectrenonics, in particular, to the concept of group time delay in quantum particles' transport in various semiconductor devices \cite{wang,ban}. One can delay or advance an incident particle or a wave packet by modulating its initial state, such as its size and incident energy and the depth of the potential well. 

\section{\label{sec:level5}Conclusion}
We solved for the expected quantum traversal time of an incident wave packet across a single potential well using the theory of time of arrival (TOA)-operators. This was done by first constructing a  TOA-operator which is the Weyl quantization of the classical TOA. The expected quantum traversal time was deduced by comparing the expectation values of the potential well and free particle TOAs for the same incident wave packet. We have found an analytic expression for the expectation value of the well-quantum traversal time where the classical contributions from the positive and negative momentum components of the incident wave packet and a purely quantum mechanical contribution due to the depth of the potential well are explicitly shown.

We also showed that the quantum well traversal time leads correctly to the barrier traversal time found in Ref. \cite{galapon1} under the variation $V_0 \to -V_0$ in the complex plane. In particular, we found that the barrier traversal time can actually be written as the sum involving the classical and quantum contributions, similar to the well traversal time. However, the presumably third term in $R_B$ with weight $\mbox{Im}\left[2\tilde{\Psi}(ik)\tilde{\Psi}^*(-ik)\right]$, which we identified earlier as a purely quantum contribution to the traversal time, exactly cancels the nonclassical contribution of the supposedly finite quantum tunneling time (that is, when $\tilde{\Psi}(\pm k)$ with $|k|>\kappa$). Hence, we only have above-the-barrier traversal time. This cancellation may physically originates from the interaction of the wavepacket components, such as interference and multiple reflections, happening inside the potential barrier. 

We also investigated the well traversal time for an incident Gaussian wave packet. Specifically, we determined how the quantum traversal time is affected by the initial state of the wave packet, such at its size ($\sigma$) and incident energy (described by the wave number $k_0=\sqrt{2\mu E_0}/\hbar$), and the depth of the potential well (described by the wave number $\kappa=\sqrt{2\mu V_0}/\hbar$). 

For shallows wells, that is, $\kappa/k_0\to 0$, we have considered two interesting cases. When $\sigma \to \infty$, the quantum well traversal time across the well is just the known classical well traversal time plus some quantum corrections mainly due to $\sigma$. This result implies that the size of the incident wave packet determines the nonclassicality of the expected traversal time across the well over mass and incident energy. On the other hand, when $\sigma \to 0$, the quantum well traversal time is just the known quantum free traversal time for free particle plus some quan tum corrections mainly due to $\kappa$, i.e., depth of the shallow potential well. This result suggests that a shallow potential acts as a small perturbation to our free particle system and induces corrections to its traversal time. 

For the case of deep potential wells, that is $\kappa/k_0 \to \infty$, we have found out that the quantum traversal time can be positive or negative implying that the particle or wave packet may be advanced or delayed for some values of $\sigma$, $\kappa$, and $k_0$. This result is different to what is classically known where the particle always speeds up at the well region and so its traversal time should decrease instead. As a possible extension of this result, it would then be interesting to investigate deeply the possible connection of the negative traversal time result we have found here to the negative phase time result found by Li and Wang \cite{Li} for future works.

\section{\label{sec:level7}Acknowledgment}
D.A.L.P. gratefully acknowledges the financial assistance from DOST-SEI through ASTHRDP-NSC graduate scholarship program. 

\appendix
\section*{Appendix}
\renewcommand\thesubsection{\Alph{subsection}}
\renewcommand{\theequation}{\Alph{subsection}.\arabic{equation}}

\subsection{Derivation of the index of refraction $R$ of the potential well}\label{app2}
\setcounter{equation}{0}
Here, we present the derivation of the effective index of refraction $R$ of the potential well given in Eq. (\ref{25th}). Recall that $R$ takes the form $R^*=k_0\int_{0}^{\infty}d\zeta \, I_0(\kappa\zeta) I(\zeta)$ where 
\begin{equation}\label{b.1}
I(\zeta)=\int_{-\infty}^{\infty}dk \, |\tilde{\Psi}(k)|^2 e^{i k \zeta}
\end{equation}
\begin{figure}
	\includegraphics[width=0.38\textwidth]{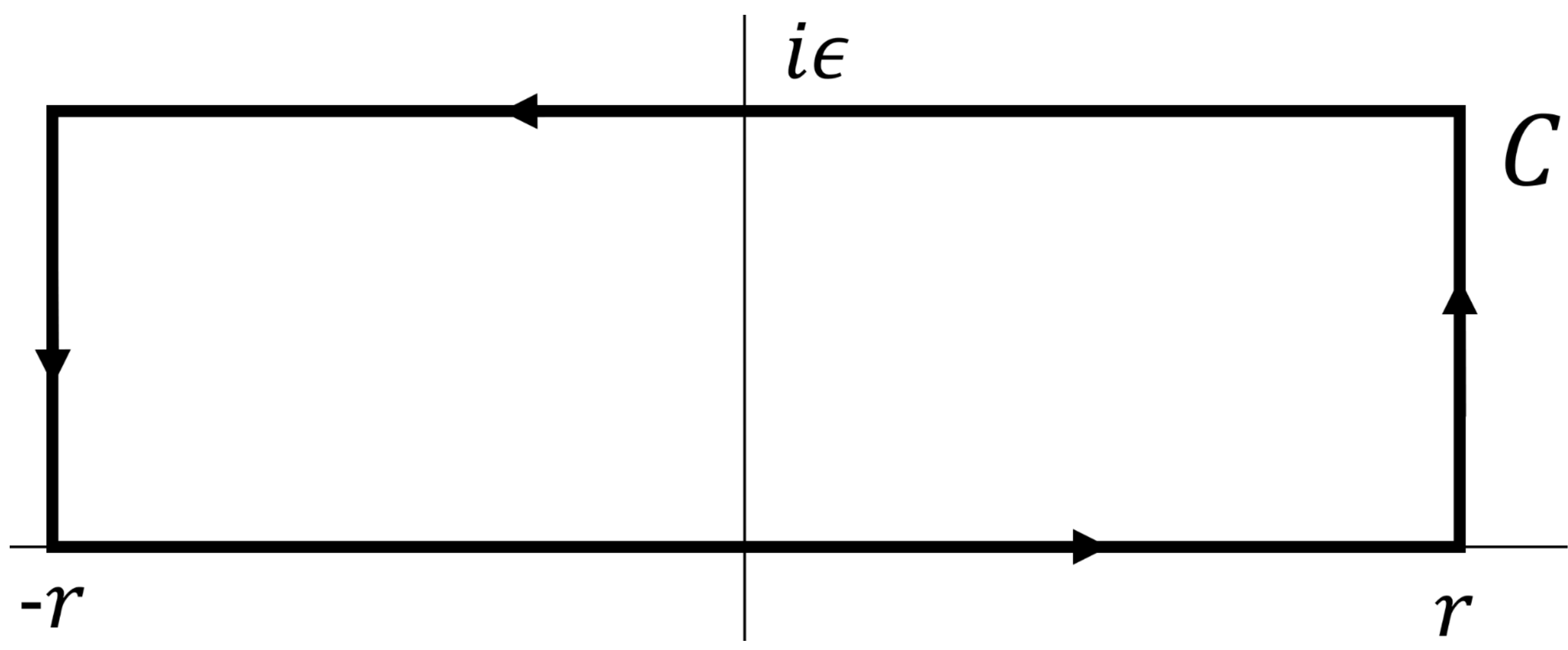}
	\caption{The contour of integration for Eq. (\ref{b.1}).}
	\label{fig:contour1}
\end{figure}
Equation (\ref{21st}) is evaluated by considering the integral $\int_C \, dz \, p(z) e^{i z \zeta}$ where $p(z)=|\tilde{\Psi}(z)|^2$ in the complex plane along the contour in Figure (\ref{fig:contour1}). For completeness, we assume that $|\tilde{\Psi}(z)|^2$ contains poles of order $n$. In the limit $r \to \infty$, we find
\begin{equation}\label{b.2}
\begin{split}
\int_{-\infty}^{\infty} dx \, p(x) e^{ix \zeta}&=\int_{-\infty}^{\infty} dx \, p(x+i\epsilon) e^{i (x+i\epsilon) \zeta} \\
& + 2 \pi i\sum_C  \mbox{Res} [p(z)]
\end{split}
\end{equation}
where $C$ indicates the contour in Fig. (\ref{fig:contour1}) enclosing the poles of $p(z)$. Using Eq. (\ref{b.2}) and interchanging the order of integration leads to 
\begin{equation}\label{b.3}
\begin{split}
R^*&=k_0\int_{-\infty}^{\infty} dk \, p (k+i\epsilon) \int_{0}^{\infty}d\zeta \, I_0(\kappa\zeta)e^{i (k+i\epsilon) \zeta}\\
& + 2 \pi i k_0  \int_{0}^{\infty}d\zeta \, I_0(\kappa\zeta)\sum  \mbox{Res} [p(z)]. 
\end{split}
\end{equation}
The interchange is valid only when $\epsilon>\kappa$ wherein the integral converges. The integral along $\zeta$ in the first term of Eq. (\ref{b.3}) evaluates to $i \thinspace \mbox{csgn}(k)/\sqrt{(k+i\epsilon)^2+\kappa^2}$, where \mbox{csgn}(k) is the cosign function, as shown. 

\begin{figure}
	\includegraphics[width=0.35\textwidth]{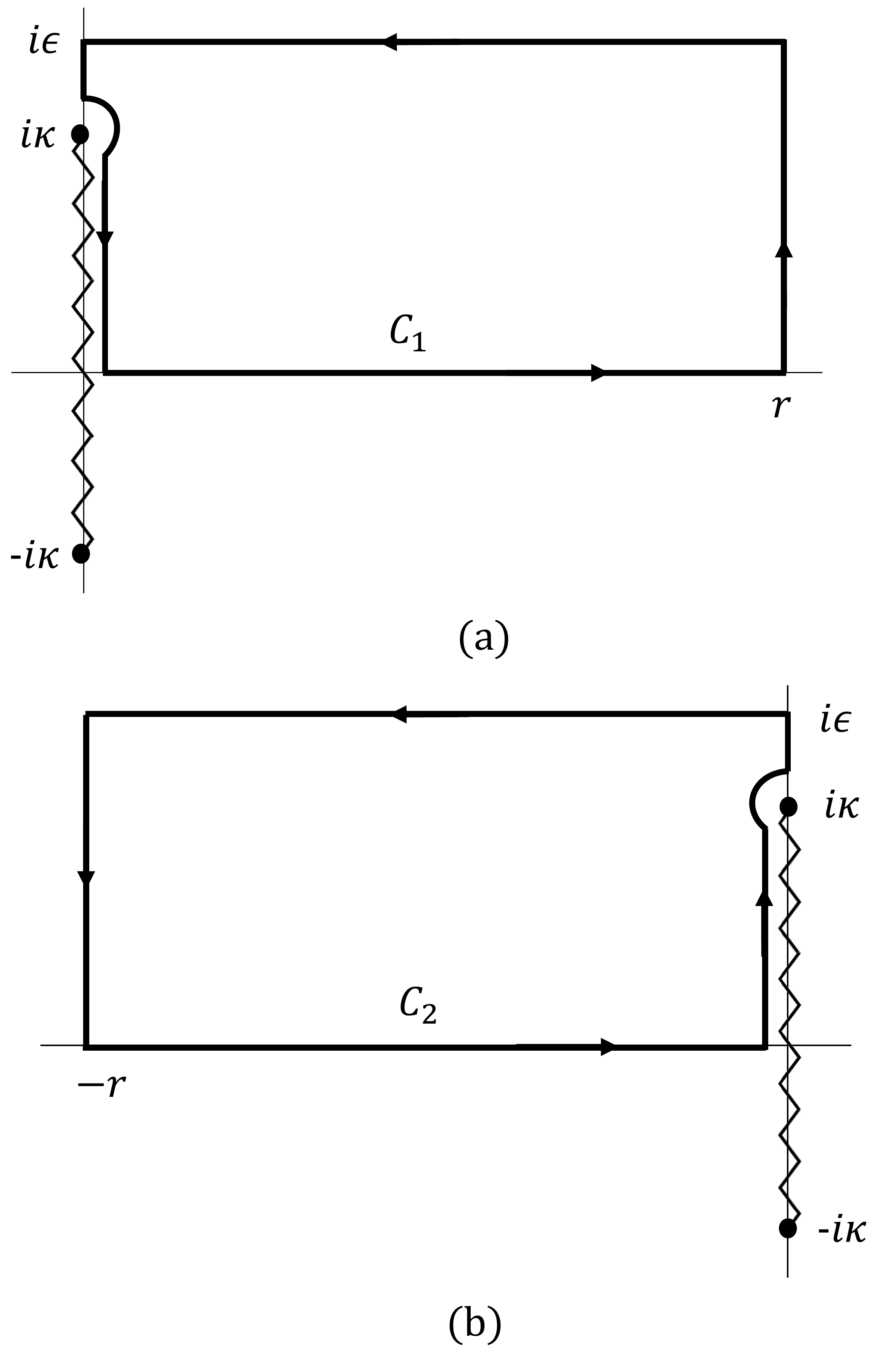}
	\caption{The contours of integration for the contour integral  $\int_C dz \, p(z)/\sqrt{z^2 + \kappa^2}.$}
	\label{fig:contour2}
\end{figure}

Consider the integral given by
\begin{equation}\label{b.4}
I^*(k)= \int_{0}^{\infty}d\zeta \, I_0(\kappa\zeta)e^{i (k+i\epsilon) \zeta}.
\end{equation}
We expand the modified Bessel function in Taylor series in Eq. (\ref{b.4}) and interchange the order of integration and summation. This leads to 
\begin{equation}\label{b.5}
I^*(k)=  \sum_{n=0}^{\infty}\bigg(\frac{\kappa}{2}\bigg)^{2n}\frac{1}{(n!)^2}\int_{0}^{\infty}d\zeta \, \zeta^{2n} e^{-\epsilon \zeta + ik \zeta}
\end{equation}
Using the integral identity $\int_{0}^{\infty}dx \,x^n e^{-\mu x}=n! \mu^{-(n+1)}$ for $Re (\mu) >0$, we find
\begin{equation}\label{b.6}
I^*(k)=\frac{i}{k+i\epsilon}\sum_{n=0}^{\infty}(-1)^n\bigg(\frac{\kappa}{2}\bigg)^{2n}\frac{(2n)!}{(n!)^2}\frac{1}{(k+i\epsilon)^{2n}}.
\end{equation}
Summing the resulting series leads to 
\begin{equation}\label{b.7}
\begin{split}
I^*(k)=\frac{i \, \mbox{csgn}(k)}{\sqrt{(k+i\epsilon)^{2} + \kappa^2}}.
\end{split}
\end{equation}
Substituting Eq. (\ref{b.7}) to Eq. (\ref{b.3}) leads to
\begin{equation}\label{b.8}
\begin{split}
R^*&=ik_0\int_{-\infty}^{\infty} dk \, \frac{p(k+i\epsilon) \mbox{csgn}(k)}{\sqrt{(k+i\epsilon)^2+\kappa^2}}\\
& + 2 \pi i k_0  \int_{0}^{\infty}d\zeta \, I_0(\kappa\zeta)\sum_C  \mbox{Res} [p(z)],
\end{split}
\end{equation}
where $p(k+i\epsilon)= \tilde{\Psi}(k+i\epsilon)\tilde{\Psi}(k-i\epsilon)^*$. To better understand the underlying physical contents of $R^*$, we rewrite the first term of Eq. (\ref{b.8}) by considering the integral $\int_C dz \, p(z)/\sqrt{z^2 + \kappa^2}$ twice in the complex plane along the two contours shown in Figure (\ref{fig:contour2}). The branch cut is set from $i\kappa$ to $-i\kappa$. In the limit $r \to \infty$, we find the integral
\begin{equation}\label{b.9}
\begin {split}
I(\pm k)&=\int_{-\infty}^{\infty} dk \, \frac{p(\pm k+i\epsilon)}{\sqrt{(\pm k+i\epsilon)^2+\kappa^2}}  \\
&=\int_{0}^{\infty} dk \, \frac{p(\pm k)}{\sqrt{k^2+\kappa^2}} + \int_{\epsilon}^{\kappa}i dk \, \frac{p(i k)}{\sqrt{\kappa^2-k^2}}\\
& \mp \int_{0}^{\kappa}i dk \, \frac{p(i k)}{\sqrt{\kappa^2-k^2}} - 2 \pi i  \sum_{C\pm} \mbox{Res} \bigg[\frac{p(z)}{(z^2 + \kappa^2)^{1/2}} \bigg],
\end {split}
\end{equation}
where $I(\pm k)$ is evaluated using the contour $C\pm$. Using Eq. (\ref{b.9}), we can rewrite Eq. (\ref{b.8}) as
\begin{equation}\label{b.10}
\begin{split}
R^*=&ik_0\int_{0}^{\infty} dk \,\frac{|\tilde{\Psi}(k)|^2}{\sqrt{k^2+\kappa^2}} - ik_0\int_{0}^{\infty}dk \,\frac{|\tilde{\Psi}(-k)|^2}{\sqrt{k^2+\kappa^2}}\\
&-2ik_0Im\bigg[\int_{0}^{+\kappa}dk \, \frac{\tilde{\Psi}(ik)\tilde{\Psi}^*(-ik)}{\sqrt{\kappa^2-k^2}}\bigg] + R_{Res}^*.
\end{split}
\end{equation}
where $R_{Res}^*$ is the contribution from the residue terms which is given by
\begin{equation}\label{b.11}
\begin{split}
&R_{Res}^*=2 \pi i k_0 \bigg[ \int_{0}^{\infty}d\zeta \, I_0(\kappa\zeta)\sum_C  \mbox{Res} [p(z)]\\
&+\sum_{C+} \mbox{Res} \bigg(\frac{p(z)}{(z^2 + \kappa^2)^{1/2}} \bigg)+  \sum_{C-} \mbox{Res} \bigg(\frac{p(z)}{(z^2 + \kappa^2)^{1/2}} \bigg)\bigg]
\end{split}
\end{equation}

It turns out that the first term of Eq. (\ref{b.11}) cancels the second and third terms so that $R_{Res}^*=0$. This suggests that there is no contribution from the poles of $|\tilde{\Psi}(z)|^2$, if exists, to the effective index of refraction $R$ of the well and in turn to the expected quantum traversal time. Taking the imaginary part of Eq. (\ref{b.10}) leads to
\begin{equation}
\begin{split}
R=k_0&\int_{0}^{\infty} dk \,\frac{|\tilde{\Psi}(k)|^2}{\sqrt{k^2+\kappa^2}} - k_0\int_{0}^{\infty}dk \,\frac{|\tilde{\Psi}(-k)|^2}{\sqrt{k^2+\kappa^2}}\\
&-2k_0 \mbox{Im}\bigg[\int_{0}^{\kappa}dk \, \frac{\tilde{\Psi}(ik)\tilde{\Psi}^*(-ik)}{\sqrt{\kappa^2-k^2}}\bigg].
\end{split}
\end{equation}
which is our final expression for the effective index of refraction of the potential well.


\end{document}